\documentclass[11pt]{article} 

\usepackage[a4paper, margin=1in]{geometry} 

\usepackage[T1]{fontenc}       
\usepackage[utf8]{inputenc}    
\usepackage{lmodern}           
\usepackage{microtype}         

\usepackage{amsmath, amssymb, amsthm}  
\usepackage{mathtools}                 
\usepackage{bbm}                        
\usepackage{bm}                         
\usepackage{commath}                    
\usepackage{nicefrac}                   
\usepackage{xfrac}                      
\usepackage{amsfonts}

\usepackage{mathrsfs}        
\DeclareMathAlphabet{\mathscr}{OMS}{rsfs}{m}{n}  

\theoremstyle{plain}

\newtheorem{theorem}{Theorem}[section]
\newtheorem{lemma}[theorem]{Lemma}

\theoremstyle{definition}

\newtheorem{remark}[theorem]{Remark}

\newtheoremstyle{boldassumption}  
  {\topsep}                       
  {\topsep}                       
  {}                     
  {}                              
  {\bfseries}                     
  {.}                             
  { }                             
  {}                              

\theoremstyle{boldassumption}
\newtheorem{assumption}{Assumption}

\usepackage{graphicx}         
\usepackage{caption}          
\usepackage{subcaption}       
\usepackage{booktabs}         
\usepackage{multirow}         
\usepackage{array}            
\usepackage{float}            
\usepackage{wrapfig}          

\usepackage[authoryear]{natbib}  
\usepackage{hyperref}         
\hypersetup{
    colorlinks=true,          
    linkcolor=blue,           
    citecolor=blue,           
    urlcolor=blue             
}

\usepackage{tikz}             
\usetikzlibrary{
    decorations.pathmorphing, 
    positioning,              
    arrows,                   
    decorations.pathreplacing,
    calc                     
}

\usepackage{pgfplots}
\usepgfplotslibrary{fillbetween} 
\pgfplotsset{compat=1.17} 

\usepackage{algorithm}        
\usepackage{algpseudocode}    

\usepackage{colortbl}      
\usepackage{hhline}        
\usepackage{calc}          
\usepackage{tabularx}      
\usepackage{comment}       

\usepackage{xcolor}           
\usepackage{enumitem}         
\usepackage{xspace}           
\usepackage{fancyvrb}         
\usepackage{tcolorbox}        

\usepackage{todonotes}        

\newcommand{\E}{\mathbb{E}}           

\newcommand{\red}{\textcolor{red}}

\allowdisplaybreaks        

\usepackage{titletoc}
\titlecontents{section}
  [0em]
  {}
  {\contentslabel{2.3em}}
  {}
  {\titlerule*[0.5pc]{.}\contentspage}

\titlecontents{subsection}
  [2.3em]
  {}
  {\contentslabel{3.2em}}
  {}
  {\titlerule*[0.5pc]{.}\contentspage}

\title{Formulating Cross-World Mediation Estimands Through Single-World Mixtures} 

\usepackage{authblk}
\author{Razieh Nabi\thanks{Email: \texttt{razieh.nabi@emory.edu}} }
\author{David Benkeser\thanks{Email: \texttt{benkeser@emory.edu}}}

\affil{Department of Biostatistics and Bioinformatics, Emory University, Atlanta, GA, USA} 

\date{}

\begin{document}
\maketitle 

\begin{abstract}
In causal mediation analysis, natural direct and indirect effects are defined through nested counterfactuals that combine the outcome under one exposure level with the mediator value under another and are therefore inherently cross-world. Their canonical identification additionally relies on cross-world independence assumptions. Consequently, both the estimands and their identifying assumptions remain controversial. 
In this paper, we ask what additional single-world structure would be required to re-express cross-world estimands as single-world quantities, identifiable under single-world assumptions. We show that these quantities can be written as mixtures of controlled single-world effects under assumptions involving a \textit{susceptibility marker}, a possibly latent baseline variable that encodes the ``would-be'' mediator value under a reference treatment arm. If observed, the marker would make several implications of the marker restrictions empirically testable. These results provide a transparent single-world formulation of cross-world mediation effects while making explicit the assumptions required. Although these conditions clarify how cross-world estimands can be interpreted within a single-world framework, their practical relevance depends on whether such susceptibility markers can be justified in specific applications. We discuss implications for principal stratification and show in the Supplementary Material how the formulation extends to ordered mediators and settings with exposure-induced mediator--outcome confounding.
\end{abstract}

\section{Introduction}
\label{sec:intro}

Cross-world estimands such as the natural direct, indirect, and path-specific effects play a central role in causal mediation analysis, yet their conventional nonparametric identification typically relies on cross-world independence assumptions relating counterfactuals from incompatible worlds \citep{robins1992identifiability,pearl2001direct,chen05ijcai}. Such independencies are implied by multiple-world causal models such as the nonparametric structural equations model with independent errors (NPSEM-IE) \citep{pearl00causality,robins2010alternative}, but not by single-world formulations such as the finest fully randomized causally interpretable structured tree graph (FFRCISTG) model \citep{robins86new,thomas13swig}. Moreover, these assumptions are empirically unverifiable and can be difficult to justify from substantive background knowledge \citep{andrews2021insights}. 

In settings where cross-world estimands are used, different estimands that avoid cross-world counterfactuals have been advocated. In particular, a substantial literature has developed and examined interventional (randomized, stochastic, or organic) effects \citep{vanderweele2014effect,vansteelandt2017interventional,rudolph2018robust,diaz2020causal,miles2023causal,lok2021causal} and separable effects \citep{stensrud2021generalized,stensrud2022separable}. Others have pursued partial identification of natural cross-world effects \citep{tchetgen2014bounds,miles2017partial}. Previous work has also connected classical cross-world estimands to effects defined through intervention-based or separable treatment-component formulations, with equality obtained under additional structural conditions \citep{didelez2006direct,robins2010alternative,robins2022interventionist}. Our goal is different: we describe conditions under which classical cross-world estimands can be represented as mixtures of controlled, single-world counterfactual means indexed by a baseline susceptibility marker, and then state additional single-world conditions under which the mixtures reduce to observed-data functionals. The construction is related to ideas from the response-type and principal-stratification literature \citep{angrist1996identification,frangakis2002principal}, but here the marker is used to clarify the single-world mixture representation of cross-world estimands, its interpretation, and the additional conditions required for identification. We first briefly review the standard mediation context and the identification assumptions commonly invoked for natural and path-specific effects.

\subsection{Cross-world mediation parameters}  

Suppose we have access to iid data on the random vector $O=(X, A, M, Y) \sim P,$ where $X$ is a vector of baseline covariates, $A \in \{0, 1\}$ the binary exposure, $M$ the mediator, and $Y$ the outcome of interest, and $P$ is a probability distribution assumed to lie in a specified statistical model. We first review the classical cross-world definitions and identification assumptions of direct and indirect effects, then review alternative formulations that allow investigators to sidestep such controversial independencies while still probing pathways.

Let $Y(a)$ denote the potential outcome under exposure $a$, and let $Y(a,m)$ denote the potential outcome if the exposure were set to $a$ and the mediator to $m$. Let $M(a)$ denote the potential mediator value under exposure $a$. Under the composition property \citep{chen05ijcai}, $Y(a)=Y(a,M(a))$, meaning that $Y(a,M(a))$ represents the outcome under $A=a$ with the mediator taking its natural value under the same exposure level. The average causal effect (ACE) is defined as $\E[Y(1) - Y(0)]$. 

Let the nested counterfactual $Y(a,M(a'))$ denote the potential outcome that would be observed if the exposure were set to $A=a$ while the mediator were set to the value it would have taken under exposure $A=a'$. The total effect admits two natural-effect decompositions:
\begin{align}
\E[Y(1)-Y(0)]
&=
\underbrace{\E[Y(1,M(0))-Y(0,M(0))]}_{\text{pure natural direct effect}}
+
\underbrace{\E[Y(1,M(1))-Y(1,M(0))]}_{\text{total natural indirect effect}}
\notag\\
&=
\underbrace{\E[Y(1,M(1))-Y(0,M(1))]}_{\text{total natural direct effect}}
+
\underbrace{\E[Y(0,M(1))-Y(0,M(0))]}_{\text{pure natural indirect effect}}.
\label{eq:natural_effect_decompositions}
\end{align}
The first decomposition pairs the pure natural direct effect with the total natural indirect effect, whereas the second pairs the total natural direct effect with the pure natural indirect effect. The four effects are defined by their respective displayed contrasts in \eqref{eq:natural_effect_decompositions}. The two decompositions coincide componentwise in the absence of treatment--mediator interaction on the additive scale, but they generally yield different direct and indirect components otherwise. 

Both decompositions above involve nested counterfactuals of the form $Y(a,M(a'))$. A sufficient set of assumptions to identify $\E[Y(a,M(a'))]$, and hence each of the natural direct and indirect effects in \eqref{eq:natural_effect_decompositions}, as a functional of the observed-data distribution $P$ is:

\begin{assumption}\label{ass:consistency}
    \emph{Causal consistency:} If $A = a$, then $M = M(a)$. If additionally $M = m$, then $Y = Y(a, m)$.
\end{assumption} 

\begin{assumption}\label{ass:positivity}
    \emph{Positivity:} $P(A = a \mid X=x) > 0$ and $P(M = m \mid A=a, X=x) > 0$, for all possible levels of $a, m$, and all $x$ such that $P(X=x) > 0$.  
\end{assumption}

\begin{assumption}\label{ass:single_world_ignor}
    \emph{Single-world ignorability:} $\{Y(a,m), M(a)\} \perp A\mid X$ and $Y(a,m)\perp M\mid A=a,X$, for all $a,m$.
\end{assumption}

\begin{assumption}\label{ass:cross_world}
    \emph{Cross‐world ignorability:} 
    \begin{align}
     Y(a,m) \perp M(a') \mid X \ , \quad \forall m, a, a', a \not= a' \ . 
    \label{eq:cross_world_assump}
    \end{align}
\end{assumption}
\noindent 
Under Assumptions~\ref{ass:consistency}-\ref{ass:cross_world}, $\E[Y(a, M(a'))]$ is identified as follows: 
\begin{align}
    \E[Y(a, M(a'))] = \int y \, dP(y \mid a, m, x)\, dP(m \mid a', x) \, dP(x) \ . 
    \label{eq:id}
\end{align}

Among the ignorability assumptions, Assumption~\ref{ass:cross_world} is the most controversial, as it entails independence relations across incompatible hypothetical worlds. To better understand the controversy underlying such assumptions, we next contrast two major causal modeling frameworks, the NPSEM-IE and FFRCISTG causal models, and examine how each treats cross-world independencies.

\begin{figure}[t] 
	\begin{center}
    \scalebox{0.55}{
    \begin{tikzpicture}[>=stealth, node distance=2cm]
        \tikzstyle{format} = [thick, circle, minimum size=1.0mm, inner sep=2pt]
        \tikzstyle{square} = [draw, thick, minimum size=4.5mm, inner sep=2pt]

    \begin{scope}[xshift=0cm, yshift=4.cm]
		\path[->, thick]
		
		node[] (a) {$A$}
		node[right of=a, xshift=0.5cm] (m) {$M$}
        node[right of=m, xshift=0.5cm] (y) {$Y$}
        node[above=0.4in of m, xshift=0.0in] (x) {$X$}

        (x) edge[black] (a)
        (x) edge[black] (m) 
        (x) edge[black] (y)
		(a) edge[black] (m) 
		(m) edge[black] (y)
        (a) edge[black, out = -30, in = 190] (y)
        
        node[below=of a, xshift=1in, yshift=0.6in] (t1) {(a)} ;
		
	\end{scope}
    
    \begin{scope}[xshift=7cm, yshift=0cm]
		
		\node[] (a) {$A$};
		\node[right=0.75in of a] (m) {$M$};
        \node[right=0.75in of m] (y) {$Y$};

        \node[above=1.5in of m, xshift=0.0in] (Ma1) {$M(1)$};
        \node[above=0.6in of m, xshift=0.0in] (Ma0) {$M(0)$};

        \node[above=2.05in of m, xshift=2.15in] (Ya1m1) {$Y(1,1)$};
        \node[above=1.3in of m, xshift=2.15in] (Ya1m0) {$Y(1,0)$};
        \node[above=0.55in of m, xshift=2.15in] (Ya0m0) {$Y(0,0)$};
        \node[above=-0.2in of m, xshift=2.15in] (Ya0m1) {$Y(0,1)$}; 
        
        \path[->, thick]

        (a) edge[black] (m)
        (m) edge[black] (y)
        (a) edge[black, out = -30, in = 190] (y)

        (Ma1) edge[red, <->] (Ma0)
        (Ma1) edge[black, bend right] (m)
        (Ma0) edge[black] (m) 
         
        (Ya1m1) edge[red, <->] (Ya1m0)
        (Ya1m1) edge[red, <->, bend left=50] (Ya0m0)
        (Ya1m1) edge[red, <->, bend left=50] (Ya0m1)
        (Ya1m0) edge[red, <->, bend left=0] (Ya0m0)
        (Ya1m0) edge[red, <->, bend left=50] (Ya0m1)
        (Ya0m0) edge[red, <->, bend left=0] (Ya0m1)

        (Ya1m1) edge[black] (y)
        (Ya1m0) edge[black] (y)
        (Ya0m1) edge[black] (y)
        (Ya0m0) edge[black] (y)

        (Ma1) edge[red, <->] (Ya0m1)
        (Ma1) edge[red, <->] (Ya0m0)

        (Ma0) edge[red, <->] (Ya1m1)
        (Ma0) edge[red, <->] (Ya1m0)
        ;      
        \node[below=of a, xshift=2.in, yshift=0.5in] () {(b)} ;

	\end{scope}

    \begin{scope}[xshift=18cm, yshift=0cm]
		
		\node[] (a) {$A$};
		\node[right=0.75in of a] (m) {$M$};
        \node[right=0.75in of m] (y) {$Y$};

        \node[above=1.5in of m, xshift=0.0in] (Ma1) {$M(1)$};
        \node[above=0.6in of m, xshift=0.0in] (Ma0) {$M(0)$};

        \node[above=2.05in of m, xshift=2.15in] (Ya1m1) {$Y(1,1)$};
        \node[above=1.3in of m, xshift=2.15in] (Ya1m0) {$Y(1,0)$};
        \node[above=0.55in of m, xshift=2.15in] (Ya0m0) {$Y(0,0)$};
        \node[above=-0.2in of m, xshift=2.15in] (Ya0m1) {$Y(0,1)$}; 
        
        \path[->, thick]

        (a) edge[black] (m)
        (m) edge[black] (y)
        (a) edge[black, out = -30, in = 190] (y)

        (Ma1) edge[red, <->] (Ma0)
        (Ma1) edge[black, bend right] (m)
        (Ma0) edge[black] (m) 
         
        (Ya1m1) edge[red, <->] (Ya1m0)
        (Ya1m1) edge[red, <->, bend left=50] (Ya0m0)
        (Ya1m1) edge[red, <->, bend left=50] (Ya0m1)
        (Ya1m0) edge[red, <->, bend left=0] (Ya0m0)
        (Ya1m0) edge[red, <->, bend left=50] (Ya0m1)
        (Ya0m0) edge[red, <->, bend left=0] (Ya0m1)

        (Ya1m1) edge[black] (y)
        (Ya1m0) edge[black] (y)
        (Ya0m1) edge[black] (y)
        (Ya0m0) edge[black] (y)
        ;      
        \node[below=of a, xshift=2.in, yshift=0.5in] () {(c)} ;

	\end{scope}
    \begin{scope}[xshift=0cm, yshift=0cm]
		\path[->, thick]
		
		node[] (a) {$A$}
		node[right of=a, xshift=0.5cm] (m) {$M$}
        node[right of=m, xshift=0.5cm] (y) {$Y$}
        node[above right of=m, yshift=0.25cm] (x) {$X$}
        node[left of=x, xshift=-0.75cm] (s) {\red{$S_M$}}

        (x) edge[black] (a) 
        (x) edge[black] (m) 
        (x) edge[black, dashed, -] (s)
        (x) edge[black] (y)
		(a) edge[black] (m) 
		(m) edge[black] (y)
        (s) edge[black] (m)
        (a) edge[black, out = -30, in = 190] (y)
        
        node[below of=m, xshift=0cm, yshift=0.7cm] () {(d)} ;
		
	\end{scope}
    \end{tikzpicture}
    }
    \end{center}
    \vspace{-0.15in}
\caption{(a) Mediation graph with treatment $A$, mediator $M$, outcome $Y$, and confounders $X$; (b, c) Causal diagrams illustrating the relationships among observed and counterfactual variables under (b) the FFRCISTG model \citep{robins86new} and (c) the NPSEM-IE model \citep{pearl00causality}, with confounders $X$ omitted for clarity; and (d) the mediation graph in (a), augmented with a candidate mediator susceptibility marker $S_M$. Red bidirected arrows in (b) and (c) indicate dependencies among counterfactual variables permitted under each model. The dashed edge in (d) denotes possible marginal dependence between $S_M$ and $X$, which need not be causal.}
\label{fig:masterdag}
\end{figure}
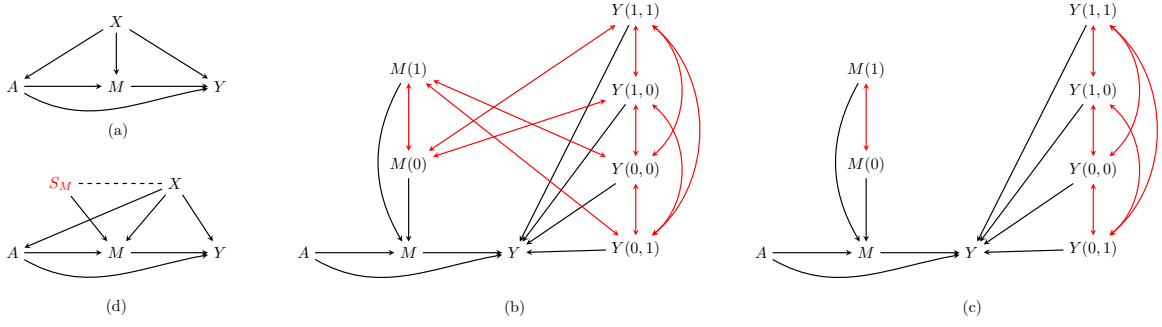

\subsection{The NPSEM-IE versus FFRCISTG causal models}

Figure~\ref{fig:masterdag} contrasts the basic mediation structure in panel (a) with the counterfactual dependencies implied by the FFRCISTG and NPSEM-IE models in panels (b) and (c), respectively. 

Under Pearl's NPSEM-IE model \citep{pearl00causality}, the structural equations are assumed to have mutually independent error terms. This independence induces the cross-world counterfactual independencies used for identification of expressions such as $\E[Y(a, M(a'))]$. 

By contrast, under Robins' FFRCISTG model \citep{robins86new, robins2010alternative}, counterfactual independence restrictions are specified within a single intervention world. The model may restrict joint distributions such as those involving $\{Y(a,m),M(a)\}$, but it does not itself impose independence between $Y(a,m)$ and $M(a')$ for $a\neq a'$. Thus, cross-world independence relations are not supplied by the FFRCISTG model.

Graphically, this distinction appears in Figures~\ref{fig:masterdag}(b, c), excluding measured confounders $X$ for simplicity. In the FFRCISTG representation (panel b), red bidirected edges connect potential outcomes across different treatment levels, reflecting that independence between $Y(a,m)$ and $M(a')$ for $a\neq a'$ is not imposed. In contrast, under the NPSEM-IE (panel c), these bidirected edges are absent, reflecting the model-implied independencies across potential outcomes indexed by different treatment levels; i.e., $Y(a,m)\perp M(a')$ for all $a,a' \in \{0,1\}$. This tension has led to several alternative mediation formulations, which we briefly review next. 

\subsection{Existing mediation formulations with no cross-world assumption} 

Cross-world assumptions have long been criticized for being empirically unverifiable and conceptually opaque \citep{robins2010alternative, andrews2021insights}. A large body of work now offers different estimands. Some replace natural effects with interventional (randomized, stochastic, or organic) effects that do not require cross-world  counterfactuals \citep{vanderweele2014effect, vansteelandt2017interventional, rudolph2018robust, diaz2020causal, miles2023causal, lok2021causal}. Another line of work argues that many questions motivating natural effects are better represented as separable effects, which consider modified versions or components of treatment so that direct and indirect effects can be defined within a single world \citep{robins2010alternative, robins2022interventionist, stensrud2021generalized, stensrud2022separable}. Natural effects have also been partially identified through bounds under weaker assumptions \citep{robins2010alternative, tchetgen2014bounds, miles2017partial}. 

\textit{Interventional (randomized, stochastic, or organic) effects} redefine the mediation estimand by replacing nested counterfactuals with contrasts between feasible intervention regimes on the mediator. Such effects remain within a single potential‐outcome world by substituting the unobservable mediator value $M(a')$ with an independent draw of a random variable $M^*(a')\sim P(M(a'))$. The \textit{interventional indirect effect} is defined as $\E[Y(0,M^*(1)) - Y(0,M^*(0))],$ and the \textit{interventional direct effect} as $\E[Y(1,M^*(1)) - Y(0,M^*(1))]$, with the sum of the two effects recovering the total causal effect of $A$ on $Y$ under the intervention that sets $A = 1$ and draws $M^*(1)\sim P(M(1))$ versus one that sets $ A = 0$ and draws $M^*(0) \sim P(M(0))$. Identification requires only Assumptions \ref{ass:consistency}-\ref{ass:single_world_ignor} above. The main trade‐off with interventional effects is interpretability: instead of asking ``how much of the treatment effect is mediated through $M$?'', it addresses the policy question ``what if the mediator's distribution under $A=0$ were replaced by that under $A=1$?''. The magnitude of this estimand depends on the intervention distribution, and does not necessarily represent a mechanism of action of the treatment. Arguably, it does not correspond to any intervention that would plausibly be implemented either. For example, doctors will rarely implement such engineered, random regimes.

\textit{Separable-effects} are defined with an interventionist intention \citep{robins2010alternative}, concerning the effect of new, modified treatments. To fix ideas, consider the special setting where the treatment consists of two distinct components, $A=(A_{M},A_{Y})$, and let $Y(a_{M},a_{Y})$ denote the potential outcome under $A_M = a_M$ and $A_Y = a_Y$. Informally, we invoke the structural assumption that $A_{M}$ influences $Y$ only through $M$, and $A_{Y}$ influences $Y$ only outside $M$. The \emph{separable indirect effect} is then defined as $\E[Y(1, 1) - Y(0, 1)]$, and the \emph{separable direct effect} as $\E[Y(0, 1) - Y(0,0)]$,  where the sum again equals the total effect. Identification requires additional dismissible-component conditions, but not cross‐world counterfactual independence \citep{stensrud2021generalized,stensrud2022separable}. Articulating a concrete decomposition story forces the investigator to clarify the scientific aim of the analysis. As a consequence, the practical relevance of the estimand becomes more transparent: is the effect of the decomposed treatments of practical interest? If this story is difficult to relate to a plausible intervention, the relevance is less clear. This feature is often viewed as desirable compared to the conceptual shortcut of vaguely defining interventions on mediators \citep{robins2010alternative,robins2022interventionist,stensrud2022separable,stensrud2023conditional}. However, the credibility of the story is context dependent: it may be reasonable in carefully engineered settings (e.g., vaccine trials with immunogenic versus reactogenic constituents), but more contentious for other interventions \citep{robins2022interventionist,young2021identified}. Yet, for these other interventions, the interpretation of classical cross-world estimands can also be difficult.

\textit{Partial identification} approaches derive ranges of effect values compatible with the observed data under assumptions weaker than those required for point identification \citep{robins2010alternative,tchetgen2014bounds,miles2017partial}. For binary treatment, mediator, and outcome, Fréchet-type arguments provide valid observable-data bounds on natural direct and indirect effects when cross-world independence is not imposed \citep{robins2010alternative,tchetgen2014bounds}; these bounds need not be sharp \citep{breum2025bounds}. Extensions accommodate polytomous mediators \citep{miles2017partial}, while response-function formulations permit linear-programming derivations of sharp bounds for a characterized class of finite discrete causal models \citep{sachs2023general}. Such bounds may nevertheless be wide \citep{miles2017partial}, although substantively justified structural or monotonicity restrictions can narrow them \citep{sjolander2009bounds,chiba2013alternative}. 

In the next section, we introduce an alternative set of assumptions under which cross-world estimands can be expressed as mixtures of single-world contrasts. We do not claim that these assumptions are plausible in most settings. However, when they do hold, they can clarify interpretation by mapping a cross-world estimand to a single-world alternative. 

\section{Single-world reinterpretations of direct and indirect effects}
\label{sec:single_mediator}

\subsection{Single-world susceptibility marker}  

To develop the single-world representations below, we introduce a candidate binary susceptibility marker $S_M$ for the binary mediator $M$. We interpret $S_M=1$ as the presence of susceptibility to mediator activation and $S_M=0$ as its absence. As depicted in Figure~\ref{fig:masterdag}(d), $S_M$ is a baseline variable that may be associated with $X$ and may affect whether $M$ becomes active, but it neither affects nor is affected by treatment assignment. We consider the following necessity condition and two alternative sufficiency conditions.

\begin{assumption}\label{ass:necessity}
    \textit{Necessity}: For each $b\in\{0,1\}$, $P(M(b)=1\mid S_M=0,X)=0.$  
\end{assumption}

\begin{assumption}\label{ass:suff_control}
    \textit{Sufficiency-under-control:} $P(M(0)=1\mid S_M=1,X)=1.$ 
\end{assumption}

\begin{assumption}\label{ass:suff_trt}
    \textit{Sufficiency-under-treatment}: $P(M(1)=1\mid S_M=1,X)=1.$ 
\end{assumption} 

Assumption~\ref{ass:necessity} states that the susceptibility marker is necessary for the mediator to be active, irrespective of the treatment status. Assumption~\ref{ass:suff_control} states that, in the absence of treatment, the presence of the susceptibility marker guarantees an active mediator. Assumption~\ref{ass:suff_trt} states that, under treatment, the presence of the susceptibility marker guarantees an active mediator. Which form of sufficiency is relevant depends on whether the cross-world parameter of interest involves $M(0)$ or $M(1)$. 

\begin{lemma}\label{lem:sufficiency}
Together Assumptions~\ref{ass:necessity} and \ref{ass:suff_control} imply
\begin{align*}
    S_M = M(0) \ .  
\end{align*}
Furthermore, together Assumptions~\ref{ass:necessity} and \ref{ass:suff_trt} imply 
\begin{align*}
    S_M = M(1) \ . 
\end{align*}
\end{lemma}%
See Appendix~\ref{app:proofs_sufficiency} for a proof. 

Lemma~\ref{lem:sufficiency} provides the basis for replacing the relevant counterfactual mediator with the susceptibility marker in the representations developed below. Importantly, $S_M$ is not defined to equal $M(a')$ for a reference treatment level $a'\in\{0,1\}$; the relevant equality follows from the marker restrictions and is not a tautology. The susceptibility marker should be understood as a structural baseline variable that summarizes whether the mediator would occur under the reference treatment. It need not be observed for the representation itself; whether it is observed becomes relevant when identification from observed data is considered.

\begin{remark}
\textbf{(Monotonicity implications).} 
Assumption~\ref{ass:necessity}, together with either Assumption~\ref{ass:suff_control} or~\ref{ass:suff_trt}, implies a one-sided monotonicity condition on the mediator's potential outcomes. Under sufficiency-under-control, $M(0)=1$ whenever $S_M=1$, while necessity implies $M(1)=0$ whenever $S_M=0$. Therefore, $M(1)\leq M(0)$: treatment may prevent mediator activation or leave the mediator unchanged, but cannot induce it. Under sufficiency-under-treatment, $M(1)=1$ whenever $S_M=1$, while necessity implies $M(0)=0$ whenever $S_M=0$. Therefore, $M(0)\leq M(1)$: treatment may induce mediator activation or leave the mediator unchanged, but cannot prevent it. Thus, the two forms of sufficiency correspond to opposite monotonicity restrictions on the pair $(M(0),M(1))$. If $S_M$ were observed and consistency and adequate positivity held, the necessity and sufficiency restrictions would imply empirically falsifiable restrictions on the distribution of $M$ within the treatment arms.
\end{remark}

\subsection{Examples illustrating the susceptibility marker}

We now provide concrete examples illustrating settings in which the assumptions introduced above may be plausible. The examples are organized into two matched pairs that retain the same treatment settings while changing the mediator under consideration. For the sufficiency-under-control member of each pair, $M=1$ denotes a state that treatment may prevent: infection following vaccination or fecal contamination following chlorination. For the sufficiency-under-treatment member, $M=1$ denotes a state that treatment may induce: an antibody response following vaccination or a target disinfectant residual following chlorination. The former states happen to be adverse and the latter protective, but this distinction is a feature of the chosen examples rather than of the assumptions. The paired examples therefore concern different mediators within the same scientific settings, rather than different assumptions imposed on the same mediator. 

\subsubsection{Necessity and sufficiency-under-control}
\label{subsubsec_suff_control}

Assumptions~\ref{ass:necessity} and~\ref{ass:suff_control} may hold when the active mediator state requires a pre-existing, treatment-unaffected enabling condition and occurs under control whenever that condition is present. By Lemma~\ref{lem:sufficiency}, the susceptibility marker then identifies those whose mediator would be active under control, while treatment may prevent activation among some in this group. We illustrate this construction using infection following vaccination and fecal contamination following chlorination. 

\vspace{0.2cm}
\noindent \textbf{\textit{Vaccination with infection as the mediator.}} 
Consider a vaccine trial in which $A\in\{0,1\}$ indicates vaccination status (e.g., control vs.\ active vaccine), $M\in\{0,1\}$ whether infection occurs, $Y$ an adverse event of special interest (AESI), and $X$ baseline covariates (e.g.\ age, comorbidities). AESIs are prespecified medically significant events selected for close monitoring because they are of scientific or medical concern in relation to a vaccine product or vaccination program \citep{who2020aesi}. For example, myocarditis occurs rarely after mRNA COVID-19 vaccination and may also follow SARS-CoV-2 infection \citep{oster2022myocarditis,patone2022risk}. Thus, there may be interest in decomposing the total effect of vaccination on myocarditis into a component operating through infection and a component operating through other pathways, presumably related to unintended side effects of the vaccine.

Infection status can in principle be manipulated through non-vaccine interventions aimed at causing infection, such as controlled human infection models, and interventions aimed at preventing infection, including masking and distancing. Let $S_M$ be an idealized binary marker summarizing a baseline exposure--susceptibility profile that is not affected by vaccination and that, in the absence of vaccination, determines whether pathogen exposure and host susceptibility are sufficient for infection. Under necessity, individuals with $S_M=0$ remain uninfected regardless of vaccination. Under sufficiency-under-control, those with $S_M=1$ who are unvaccinated become infected almost surely. Together, these restrictions imply $S_M=M(0)$. Thus, $S_M=1$ identifies individuals who would become infected if unvaccinated, whereas $S_M=0$ identifies those who would remain uninfected under either vaccination status.

\vspace{0.2cm}
\noindent \textbf{\textit{Chlorination with fecal contamination as the mediator.}} 
Let $A \in \{0,1\}$ indicate whether a community chlorination program is implemented, with $A=1$ representing chlorination and $A=0$ no chlorination. Define the mediator $M \in \{0,1\}$ as fecal contamination at the household tap, operationalized by the presence of detectable \textit{E. coli}, a standard indicator of fecal contamination in drinking water \citep{who2022drinkingwater}. Let $Y$ denote a downstream health outcome such as gastrointestinal illness. Chlorination interventions have been shown to reduce detectable \textit{E. coli} contamination and diarrheal illness \citep{arnold2007treating,pickering2019effect}. The contamination state represented by $M$ is manipulable in principle, e.g., through controlled contamination or decontamination of the water. Let $S_M \in \{0,1\}$ represent the presence of an upstream fecal contamination source that, in the absence of chlorination, leads to detectable \textit{E. coli} at the tap. Necessity would require households without such a source ($S_M=0$) to remain uncontaminated regardless of treatment, and sufficiency-under-control would require upstream fecal contamination to produce downstream microbial detection whenever chlorination is absent. Together, these restrictions imply $S_M=M(0)$. Thus, $S_M=1$ identifies households that would have detectable \textit{E. coli} at the tap in the absence of chlorination, whereas $S_M=0$ identifies those that would lack detectable \textit{E. coli} under either treatment condition. 

\subsubsection{Necessity and sufficiency-under-treatment}
\label{subsubsec_suff_trt}

The preceding examples define $M=1$ as an adverse state that treatment may prevent. We now retain the same treatment settings but consider protective mediator states that treatment may induce. Assumptions~\ref{ass:necessity} and~\ref{ass:suff_trt} may hold when attainment of the active mediator state under treatment requires a pre-existing capacity to respond. By Lemma~\ref{lem:sufficiency}, the susceptibility marker then identifies those for whom the mediator would be active under treatment, while individuals with $S_M=0$ cannot attain the active mediator state under either treatment level. We illustrate this construction using antibody response following vaccination and disinfectant residual following chlorination. 

\vspace{0.2cm}
\noindent \textbf{\textit{Vaccination with antibody response as the mediator.}} 
Let $A\in\{0,1\}$ indicate vaccination, $M\in\{0,1\}$ the presence, at a prespecified post-treatment time, of an antibody level associated with protection for the infection under study (e.g., a neutralizing-antibody titer exceeding a prespecified threshold), and $Y$ a downstream outcome such as infection or disease severity. For many vaccines, antibodies induced by vaccination are correlates of protection \citep{plotkin2010correlates}. Unlike seroconversion, which denotes an endogenously generated response, the antibody state encoded by $M$ can, in principle, also be attained through passive administration of pathogen-specific antibodies, including antibody-containing plasma from convalescent individuals \citep{casadevall2004passive,casadevall2020convalescent}. Define $S_M$ as an idealized baseline immune-susceptibility marker, such as a sufficiently rich summary of B-cell function and immunosuppressive status; for example, B-cell-depleting treatment can substantially attenuate humoral responses to vaccination \citep{moor2021humoral}. Necessity would require individuals with $S_M=0$ to fail to attain the specified antibody state regardless of vaccination, while sufficiency-under-treatment would require every vaccinated individual with $S_M=1$ to attain it. Together, these restrictions imply $S_M=M(1)$. Thus, $S_M=1$ identifies individuals who would attain the specified antibody state if vaccinated, whereas $S_M=0$ identifies those who would fail to attain it under either vaccination status. 

\vspace{0.2cm}
\noindent \textbf{\textit{Chlorination with disinfectant residual as the mediator.}}
Let $A\in\{0,1\}$ indicate whether a community chlorination program is implemented, and let $M\in\{0,1\}$ indicate whether, at a prespecified post-intervention time, the free-chlorine residual at the household tap lies within a prespecified target range. Let $Y$ denote a subsequent health outcome such as gastrointestinal illness. A target disinfectant residual indicates that chlorine remains available after treatment and transport through the distribution system, although residual chlorine is not itself a complete measure of microbiological safety \citep{who2022drinkingwater}. The mediator can, in principle, be manipulated independently of the community program through adjustment of chlorine dosing or point-of-use chlorination \citep{arnold2007treating}. Define $S_M\in\{0,1\}$ as an idealized baseline chlorination-responsiveness marker summarizing pre-intervention characteristics of the household's source water and distribution path that determine whether the target residual can be attained, such as chlorine demand, turbidity, storage or travel time, and distribution-system integrity. Necessity would require households with $S_M=0$ to fail to attain the target residual regardless of whether the chlorination program is implemented, while sufficiency-under-treatment would require every household with $S_M=1$ to attain the target residual when the program is implemented. Together, these restrictions imply $S_M=M(1)$. Thus, $S_M=1$ identifies households that would attain the target chlorine residual under the chlorination program, whereas $S_M=0$ identifies those that would fail to attain it under either treatment condition. 

Across these examples, the substantive meaning of $S_M$ differs, but its structural role is the same: it represents a treatment-unaffected condition or capacity that determines mediator status under the relevant reference treatment. We now turn to the central algebraic implication: once the equality in Lemma~\ref{lem:sufficiency} is established, cross-world mediation contrasts can be rewritten as single-world mixtures.

\subsection{Rewriting cross-world mediation effects as single-world mixtures}

We now show that, under the stated assumptions, classical cross-world mediation contrasts can be rewritten as single-world expectations involving the susceptibility marker $S_M$. 

Depending on the cross-world term of interest, necessity (Assumption~\ref{ass:necessity}) combined with either sufficiency-under-control (Assumption~\ref{ass:suff_control}) or sufficiency-under-treatment (Assumption~\ref{ass:suff_trt}) identifies $S_M$ with $M(0)$ or $M(1)$. Although $M(a')$ is a single-world counterfactual, the nested form $Y(a, M(a'))$ references two treatment levels, $a$ and $a'$, making it a \emph{cross-world} quantity. Moreover, $M(a')$ is revealed in the factual world only for individuals with $A=a'$, whereas $S_M$ is formulated as a baseline variable, though it may remain latent. Once the marker restrictions establish $S_M=M(a')$, the cross-world nesting disappears, and the resulting expression can be written in terms of controlled single-world counterfactuals conditional on $S_M$.

To see this formally, fix a pair $(a, a')$ and suppose that the relevant sufficiency condition allows us to set $S_M = M(a')$. Then 
\begin{equation}\label{eq:Yam}
\begin{aligned}
    &\E[Y(a,M(a'))] \\
    &\hspace{1cm} \overset{def}{=} \sum_{m=0}^1 \E[Y(a, m) \mid M(a') =m] \, P(M(a')=m) \\ 
    &\hspace{1cm} =\sum_{m=0}^1 \E[Y(a, m) \mid S_M=m] \, P(S_M=m)  \\
    &\hspace{1cm} =\E[Y(a, m=1) \mid S_M=1] \, P(S_M=1) +  \E[Y(a, m=0) \mid S_M=0] \, P(S_M=0) \ . 
\end{aligned}
\end{equation}
Under the substitution $S_M=M(a')$, the cross-world counterfactual mean $\E[Y(a,M(a'))] $ becomes a two-component mixture of single-world counterfactual means stratified by the susceptibility marker $S_M$, with weights given by the marginal distribution of $S_M$. Thus the susceptibility-marker formulation reparameterizes the cross-world estimand as a mixture of controlled counterfactual outcomes defined within susceptibility-defined subpopulations. In other words, the cross-world counterfactual mean can be written as a weighted average of the controlled outcomes $Y(a,1)$ and $Y(a,0)$ evaluated within subpopulations defined by whether the mediator would take the value $1$ or $0$ under the reference treatment. Conceptually, the susceptibility marker converts the counterfactual mediator value $M(a')$ into a baseline characteristic that partitions the population according to whether the mediator pathway is available. 

Both the classical cross-world estimand and its single-world re-expression involve potential outcomes defined under interventions on the mediator. Their substantive interpretation therefore depends on whether the relevant mediator interventions can be meaningfully specified in the scientific setting. This requirement is distinct from the cross-world considerations studied here. Our goal is to clarify how, when the susceptibility-marker conditions hold, cross-world estimands can be represented using single-world quantities. 

The susceptibility-marker restrictions do not change the definitions of the natural effects. Rather, they determine which reference mediator value can be replaced by the marker and, therefore, which decomposition in \eqref{eq:natural_effect_decompositions} admits the single-world mixture representation developed here. Sufficiency-under-control aligns with the decomposition into the pure natural direct effect and the total natural indirect effect, whereas sufficiency-under-treatment aligns with the decomposition into the total natural direct effect and the pure natural indirect effect. We consider these two cases in turn. 

\subsubsection{Rewriting cross-world mediation effects with sufficiency-under-control}

The sufficiency-under-control construction corresponds to the first decomposition in \eqref{eq:natural_effect_decompositions}, which pairs the pure natural direct effect with the total natural indirect effect and is anchored at the mediator value under control.

To illustrate the effect decomposition under sufficiency-under-control, consider the infection example introduced in Section~\ref{subsubsec_suff_control}. Under necessity and sufficiency-under-control, Lemma~\ref{lem:sufficiency} permits $M(0)$ to be replaced by $S_M$. Setting $a'=0$ in \eqref{eq:Yam} therefore gives $\E[Y(a,M(0))] = \E[Y(a, 1) \mid S_M=1] \, P(S_M=1) +  \E[Y(a, 0) \mid S_M=0] \, P(S_M=0)$. Here, $\E[Y(a,1)\mid S_M=1]$ is the average potential outcome under treatment $A=a$ with infection fixed to be present, among individuals who would become infected if unvaccinated; $P(S_M=1)$ is the prevalence of this subgroup. Likewise, $\E[Y(a,0)\mid S_M=0]$ is the average potential outcome under $A=a$ with infection fixed to be absent, among individuals who would remain uninfected if unvaccinated. Under necessity, individuals with $S_M=0$ would in fact remain uninfected regardless of vaccination.

The pure natural direct effect, $\E[Y(1,M(0)) - Y(0,M(0))]$, can therefore be rewritten as    
\begin{equation}\label{eq:controlled_dir_eff_suff_under_control}
\begin{aligned}
    &\E[Y(1,M(0)) - Y(0,M(0))]  \\ 
    &\hspace{0.25cm}=\sum_{m=0}^1 \big\{\E[Y(1,m)\mid S_M=m] - \E[Y(0,m)\mid S_M=m]
    \big\}\,P(S_M=m) \\ 
    &\hspace{0.25cm}= 
    \underbrace{\big\{\E[Y(1,1) - Y(0,1) \mid S_M=1] \big\}}_{\Delta^\text{dir-c}_1} P(S_M=1) + 
    \underbrace{\big\{\E[Y(1,0) - Y(0,0) \mid S_M=0] \big\}}_{\Delta^\text{dir-c}_0} P(S_M=0) \ . 
\end{aligned}
\end{equation}%
The quantities $\Delta^\text{dir-c}_1$ and $\Delta^\text{dir-c}_0$ are \textit{conditional controlled direct effects}. The subgroup $S_M=1$ consists of individuals who would experience the mediator under the reference condition. The contrast $\Delta^\text{dir-c}_1 = \E[Y(1,1) - Y(0,1) \,|\, S_M=1]$ measures the effect of treatment when the mediator is fixed to its active state in this subgroup and therefore excludes the pathway through which treatment prevents mediator activation. By contrast, necessity implies that the mediator cannot occur among individuals with $S_M=0$ under either treatment level. Thus, $\Delta^\text{dir-c}_0 = \E[Y(1,0) - Y(0,0) \mid S_M=0]$ measures the treatment effect in a subpopulation in which treatment cannot operate through changes in mediator status. 

Equation~\eqref{eq:controlled_dir_eff_suff_under_control} shows that the the pure natural direct effect can be written as a weighted average of these controlled direct effects within susceptibility-marker strata. Specifically, it combines the controlled direct effect with the mediator fixed at $1$ among individuals with $S_M=1$ and the controlled direct effect with the mediator fixed at $0$ among individuals with $S_M=0$, weighted by the prevalence of the two strata. The result therefore gives the pure natural direct effect a single-world mixture interpretation in terms of controlled causal contrasts within distinct susceptibility subpopulations. 

In the infection example, $S_M=1$ identifies individuals whose pathogen exposure would produce infection in the absence of vaccination. The contrast $\Delta^\text{dir-c}_1$ compares the risk of myocarditis under vaccination versus no vaccination while infection is fixed to be present under both conditions. It therefore excludes the pathway through which vaccination affects myocarditis by changing whether infection occurs. Because the binary mediator records only the occurrence of infection, however, the contrast may still include effects of vaccination on the subsequent course or severity of infection and its downstream inflammatory consequences, as well as vaccine-related immune pathways that affect myocarditis without changing infection status. The mechanisms underlying myocarditis following SARS-CoV-2 infection and mRNA vaccination remain incompletely understood, although immune and inflammatory pathways have been proposed in both settings \citep{siripanthong2020recognizing,heymans2022myocarditis}. Thus, $\Delta^\text{dir-c}_1$ is a net controlled direct effect and should not be interpreted as isolating a single protective or adverse mechanism.

By contrast, $S_M=0$ identifies individuals who would remain uninfected regardless of vaccination. For this subgroup, fixing infection to be absent agrees with its natural infection status under either treatment level.  Consequently, $\Delta^\text{dir-c}_0$ is the effect of vaccination on myocarditis among individuals who would remain uninfected under either vaccination status. It therefore captures the net effect of vaccination through pathways other than infection. In the COVID-19 vaccine example, this provides a particularly direct causal safety contrast: a positive value represents an increased risk of myocarditis under vaccination among individuals who would remain uninfected under either vaccination status, whereas a negative value represents a protective effect within that subpopulation. 

The total natural indirect effect, $\E[Y(1,M(1)) - Y(1,M(0))]$, can be rewritten as 
\begin{equation}\label{eq:indir_eff_suff_under_control}
\begin{aligned}
    &\E[Y(1,M(1)) - Y(1,M(0))]  \\ 
    &\hspace{0.5cm}=\sum_{m=0}^1 \big\{\E[Y(1)\mid S_M=m] - \E[Y(1,m)\mid S_M=m]
    \big\}\,P(S_M=m) \\ 
    &\hspace{0.5cm}= 
    \big\{\E[Y(1) - Y(1,1) \mid S_M=1] \big\}  
    \,P(S_M=1) + 
    \big\{\E[Y(1) - Y(1,0) \mid S_M=0] \big\}
    \,P(S_M=0) \\ 
    &\hspace{0.5cm}= 
    \underbrace{\big\{\E[Y(1) - Y(1,1) \mid S_M=1] \big\}}_{\Delta^\text{indir-c}_1} \,P(S_M=1) \ . 
\end{aligned}
\end{equation}%
The last equality follows from necessity and composition: among individuals with $S_M=0$, $M(1)=0$, and hence $Y(1)=Y(1,M(1))=Y(1,0)$. The quantity $\Delta^\text{indir-c}_1 = \E[Y(1) - Y(1,1) \mid S_M=1]$ is a \textit{susceptibility-stratum-specific contrast} between the natural and controlled mediator regimes under treatment. It is not a controlled indirect effect in the conventional sense because the mediator is left at its natural treatment value in $Y(1)$ and fixed at $1$ only in $Y(1,1)$. Nevertheless, it has a clear mediator-intervention interpretation: with treatment held fixed at $A=1$, it compares the outcome when the mediator takes its natural value under treatment with the outcome when the mediator is instead fixed to its active value. 

Equation~\eqref{eq:indir_eff_suff_under_control} shows that the mediated component arises only within the susceptible subgroup $S_M=1$. Among individuals in the $(M(0),M(1))=(1,1)$ stratum, the natural and controlled mediator regimes under treatment coincide, so their contribution to $\Delta^\text{indir-c}_1$ is zero. Consequently, only individuals in the $(M(0),M(1))=(1,0)$ stratum, for whom treatment prevents mediator activation, can contribute to the total natural indirect effect. For them, the contrast compares the outcome under treatment with the mediator inactive to the outcome under treatment with the mediator fixed to be active. Thus, $\Delta^\text{indir-c}_1=\E[Y(1)-Y(1,1)\mid S_M=1]$ captures the consequence of allowing treatment to prevent mediator activation rather than overriding that action by fixing the mediator at its active value. 

In the infection example, $S_M=1$ identifies individuals who would become infected if unvaccinated. The contrast $\Delta^\text{indir-c}_1$ compares their risk of myocarditis under vaccination with infection allowed to take its natural post-vaccination value versus under vaccination with infection fixed to be present. For individuals who remain infected despite vaccination, these two regimes coincide and the contrast is zero. For individuals whose infection is prevented by vaccination, the contrast compares myocarditis risk in the absence versus presence of infection. Accordingly, $\Delta^\text{indir-c}_1$ captures the component of the vaccination effect on myocarditis operating through prevention of infection among individuals who would have become infected without vaccination. 

The components defined above recover the conditional total effect within each susceptibility-marker stratum. Specifically,
\begin{align*}
\Delta^\text{dir-c}_1+\Delta^\text{indir-c}_1
&=
\E\!\left[
Y(1,1)-Y(0,1)+Y(1)-Y(1,1)
\mid S_M=1
\right] \\
&=
\E[Y(1)-Y(0)\mid S_M=1] \ ,
\\[0.3em]
\Delta^\text{dir-c}_0
&=
\E[Y(1,0)-Y(0,0)\mid S_M=0] \\
&=
\E[Y(1)-Y(0)\mid S_M=0] \ .
\end{align*}
The first identity follows because $S_M=M(0)=1$ within the $S_M=1$ stratum, so composition gives $Y(0)=Y(0,1)$. The second follows from necessity and composition: within the $S_M=0$ stratum, $M(1)=M(0)=0$, so $Y(a)=Y(a,0)$ for $a\in\{0,1\}$.

Applying the law of total expectation therefore yields
\begin{align}
    \E[Y(1)-Y(0)]
    &=
    \big(\Delta^\text{dir-c}_1+\Delta^\text{indir-c}_1\big)P(S_M=1)
    +
    \Delta^\text{dir-c}_0P(S_M=0) \ .
    \label{eq:decompo_suff_under_control}
\end{align}
Equation~\eqref{eq:decompo_suff_under_control} expresses the total effect as a prevalence-weighted average of the conditional total effects in the two strata. Within the $S_M=1$ stratum, the conditional total effect is further decomposed into the controlled direct component $\Delta^\text{dir-c}_1$ and the mediated component $\Delta^\text{indir-c}_1$. By contrast, within the $S_M=0$ stratum, the mediator remains inactive and the conditional total effect is $\Delta^\text{dir-c}_0$. 

In the infection example, within the $S_M=1$ stratum, the conditional total effect is the sum of $\Delta^\text{dir-c}_1$, which compares vaccination with infection held present, and $\Delta^\text{indir-c}_1$, which captures the mediated component attributable to prevention of infection. The $S_M=0$ stratum consists of individuals who would remain uninfected regardless of vaccination. Within this stratum, the conditional total effect is $\Delta^\text{dir-c}_0$, the vaccination safety contrast among individuals who would remain uninfected under either vaccination status.  Thus, the decomposition distinguishes the effect of vaccination among individuals who would become infected without vaccination from its effect among those who would remain uninfected under either vaccination status.

\subsubsection{Rewriting cross-world mediation effects with sufficiency-under-treatment} 

The sufficiency-under-treatment construction corresponds to the second decomposition in \eqref{eq:natural_effect_decompositions}, which pairs the total natural direct effect with the pure natural indirect effect and is anchored at the mediator value under treatment.

An alternative additive decomposition of the total effect is $\E[Y(1)-Y(0)] = \E[Y(1)-Y(0,M(1))] + \E[Y(0,M(1))-Y(0)]$. We refer to the first term as the total natural direct effect and the second as the pure natural indirect effect associated with the mediator value under treatment. Because both components involve $M(1)$, sufficiency-under-treatment provides the relevant single-world representation.  

To illustrate the corresponding representations under sufficiency-under-treatment, return to the vaccine and immune-response example introduced in Section~\ref{subsubsec_suff_trt}, where $A$ denotes vaccination, $M$ indicates attainment of a prespecified antibody state associated with protection, and $Y$ is a subsequent outcome such as infection. Under necessity and sufficiency-under-treatment, Lemma~\ref{lem:sufficiency} permits $M(1)$ to be replaced by $S_M$. Setting $a'=1$ in \eqref{eq:Yam} therefore gives $\E[Y(a,M(1))] = \E[Y(a,1)\mid S_M=1] \, P(S_M=1) + \E[Y(a,0)\mid S_M=0] \, P(S_M=0)$. Here, $\E[Y(a,1)\mid S_M=1]$ is the average potential outcome under treatment $A=a$ with the specified antibody state fixed to be present, among individuals who would attain that state if vaccinated. Likewise, $\E[Y(a,0)\mid S_M=0]$ is the average potential outcome under $A=a$ with the antibody state fixed to be absent, among individuals who would fail to attain it if vaccinated. Under necessity, individuals with $S_M=0$ would in fact fail to attain the specified antibody state regardless of vaccination. The weights are the prevalences of these two susceptibility groups. 

The total natural direct effect, $\E[Y(1, M(1)) - Y(0,M(1))]$, can be rewritten as
\begin{equation}\label{eq:dir_eff_suff_under_trt}
\begin{aligned}
&\E[Y(1, M(1)) - Y(0,M(1))] \\
&\hspace{0.25cm}= \sum_{m=0}^1 \big\{\E[Y(1, m)\mid S_M=m] - \E[Y(0,m)\mid S_M=m]\big\}\,P(S_M=m) 
\\
&\hspace{0.25cm}=
\underbrace{\big\{\E[Y(1, 1) - Y(0,1)\mid S_M=1]\big\}}_{\Delta_1^\text{dir-t}} P(S_M=1)
+
\underbrace{\big\{\E[Y(1, 0) - Y(0,0)\mid S_M=0]\big\}}_{\Delta_0^\text{dir-t}} P(S_M=0) \ . 
\end{aligned}
\end{equation}
The first equality uses composition and $S_M=M(1)$: within the stratum $S_M=m$, $M(1)=m$ and hence $Y(1)=Y(1,m)$. The quantities $\Delta_1^\text{dir-t}$ and $\Delta_0^\text{dir-t}$ are \emph{conditional controlled direct effects}. The subgroup $S_M=1$ consists of individuals who would experience the mediator under treatment. The contrast $\Delta_1^\text{dir-t}=\E[Y(1,1)-Y(0,1)\mid S_M=1]$ measures the effect of treatment when the mediator is fixed to its active state in this subgroup and therefore excludes the pathway through which treatment induces mediator activation. By contrast, necessity implies that the mediator would remain inactive among individuals with $S_M=0$ under either treatment level. Thus, $\Delta_0^\text{dir-t}=\E[Y(1,0)-Y(0,0)\mid S_M=0]$ measures the treatment effect in a subpopulation in which treatment cannot operate through changes in mediator status. 

Equation~\eqref{eq:dir_eff_suff_under_trt} is the treatment-anchored analogue of \eqref{eq:controlled_dir_eff_suff_under_control}: it expresses the total natural direct effect as a prevalence-weighted average of conditional controlled direct effects within susceptibility-marker strata.


In the vaccine example, $S_M=1$ identifies individuals who would attain the specified antibody state if vaccinated. The contrast $\Delta_1^\text{dir-t}$ compares the risk of infection under vaccination versus no vaccination while the antibody state is fixed to be present under both conditions, among individuals who would attain that state if vaccinated. Under vaccination, this intervention agrees with the natural antibody state in this subgroup. Under no vaccination, the antibody state could, in principle, be attained through an intervention such as passive administration of pathogen-specific antibodies. The contrast therefore excludes the pathway through which vaccination affects infection by changing whether the specified antibody state is attained. Because the binary mediator records only whether a prespecified antibody threshold is reached, however, the contrast may still include effects of vaccination through other immune mechanisms or through aspects of the immune response not captured by this binary mediator. Thus, $\Delta_1^\text{dir-t}$ is a net controlled direct effect and should not be interpreted as isolating a single immune mechanism.

By contrast, $S_M=0$ identifies individuals who would fail to attain the specified antibody state regardless of vaccination. The contrast $\Delta_0^\text{dir-t}$ compares the risk of infection under vaccination versus no vaccination while the antibody state is fixed to be absent in this subgroup. Because fixing the antibody state to be absent agrees with its natural value under either treatment level, $\Delta_0^\text{dir-t}$ is also the conditional total effect of vaccination within this subgroup. It captures the net effect of vaccination through pathways that do not require attainment of the specified antibody state, which may include other components of the immune response.

The pure natural indirect effect, $\E[Y(0,M(1)) - Y(0)]$, can be rewritten as 
\begin{equation}\label{eq:indir_eff_suff_under_trt}
\begin{aligned}
&\E[Y(0,M(1)) - Y(0)] \\
&\hspace{0.5cm}= \sum_{m=0}^1 \big\{\E[Y(0,m)\mid S_M=m] - \E[Y(0)\mid S_M=m]\big\} \, P(S_M=m) \\
&\hspace{0.5cm}=  \big\{\E[Y(0,1) - Y(0) \mid S_M=1] \big\} \, P(S_M=1) + \big\{\E[Y(0,0) - Y(0) \mid S_M=0] \big\} \, P(S_M=0) \\ 
&\hspace{0.5cm}=
\underbrace{\big\{\E[Y(0,1) - Y(0)\mid S_M=1]\big\}}_{\Delta^\text{indir-t}_1} P(S_M=1) \ .
\end{aligned}
\end{equation} 
The last equality follows from necessity and composition: among individuals with $S_M=0$, $M(0)=0$, and hence $Y(0)=Y(0,M(0))=Y(0,0)$. The quantity $\Delta_1^\text{indir-t}=\E[Y(0,1)-Y(0)\mid S_M=1]$ is a \emph{susceptibility-stratum-specific contrast} between the controlled and natural mediator regimes under no treatment. It is not a controlled indirect effect in the conventional sense because the mediator is fixed at $1$ in $Y(0,1)$ but left at its natural untreated value in $Y(0)$. Nevertheless, it has a clear mediator-intervention interpretation: with treatment held fixed at $A=0$, it compares the outcome when the mediator is fixed to its active value with the outcome when the mediator instead takes its natural value under no treatment.

Equation~\eqref{eq:indir_eff_suff_under_trt} shows that the mediated component arises only within the susceptible subgroup $S_M=1$. Among individuals in the $(M(0),M(1))=(1,1)$ stratum, the controlled and natural mediator regimes under no treatment coincide, so their contribution to $\Delta_1^\text{indir-t}$ is zero. Consequently, only individuals in the $(M(0),M(1))=(0,1)$ stratum, for whom treatment induces mediator activation, can contribute to the pure natural indirect effect. For them, the contrast compares the outcome under no treatment with the mediator fixed to be active to the outcome under no treatment with the mediator inactive. Thus, $\Delta_1^\text{indir-t}$ captures the consequence of imposing the mediator state that treatment would induce rather than leaving the mediator at its natural untreated value. 

In the vaccine example, $\Delta_1^\text{indir-t}$ compares, among individuals who would attain the specified antibody state if vaccinated, the risk of infection under no vaccination with the antibody state fixed to be present versus under no vaccination with the antibody state allowed to take its natural value. For individuals who would attain the antibody state even without vaccination, these two regimes coincide and the contrast is zero. For individuals who would not attain it without vaccination but would attain it if vaccinated, the contrast compares infection risk under no vaccination with the antibody state present versus absent. Accordingly, $\Delta_1^\text{indir-t}$ captures the component of the vaccination effect on infection operating through vaccination-induced attainment of the specified antibody state. 

The components defined above recover the conditional total effect within each susceptibility-marker stratum. Specifically, 
\begin{align*} 
\Delta_1^\text{dir-t}+\Delta_1^\text{indir-t} 
&= \E\!\left[ Y(1,1)-Y(0,1)+Y(0,1)-Y(0) \mid S_M=1 \right] 
\\ 
&= \E[Y(1)-Y(0)\mid S_M=1] \ , 
\\[0.3em] 
\Delta_0^\text{dir-t} 
&= \E[Y(1,0)-Y(0,0)\mid S_M=0] \\ 
&= \E[Y(1)-Y(0)\mid S_M=0] \ . 
\end{align*} 
The first identity follows because $S_M=M(1)=1$ within the $S_M=1$ stratum, so composition gives $Y(1)=Y(1,1)$. The second follows from necessity and composition: within the $S_M=0$ stratum, $M(1)=M(0)=0$, so $Y(a)=Y(a,0)$ for $a\in\{0,1\}$. 

Applying the law of total expectation therefore yields 
\begin{align} 
\E[Y(1)-Y(0)] 
&= \big( \Delta_1^\text{dir-t} + \Delta_1^\text{indir-t} \big)P(S_M=1) + \Delta_0^\text{dir-t}P(S_M=0) \ . 
\label{eq:decompo_suff_under_trt} 
\end{align} 
Equation~\eqref{eq:decompo_suff_under_trt} is the treatment-anchored analogue of \eqref{eq:decompo_suff_under_control}: within the $S_M=1$ stratum, the conditional total effect decomposes into $\Delta^{\text{dir-t}}_1+\Delta^{\text{indir-t}}_1$, whereas within the $S_M=0$ stratum, it equals $\Delta^{\text{dir-t}}_0$.


In the vaccine example, within the $S_M=1$ stratum, the conditional total effect is the sum of $\Delta_1^\text{dir-t}$, which compares vaccination with no vaccination while the specified antibody state is held present, and $\Delta_1^\text{indir-t}$, which captures the mediated component attributable to vaccination-induced attainment of the antibody state. The $S_M=0$ stratum consists of individuals who would fail to attain the specified antibody state under either vaccination status. Within this stratum, the conditional total effect is $\Delta_0^\text{dir-t}$, the corresponding vaccination-effect contrast among individuals who would not attain that antibody state. Thus, the decomposition distinguishes the effect of vaccination among individuals who would attain the specified antibody state if vaccinated from its effect among those who would not.

Taken together, Equations~\eqref{eq:controlled_dir_eff_suff_under_control}--\eqref{eq:decompo_suff_under_trt} establish that the two natural-effect decompositions can be represented as prevalence-weighted mixtures of single-world causal contrasts evaluated within susceptibility-marker strata. In both cases, the direct components are conditional controlled direct effects, whereas the mediated components are susceptibility-stratum-specific contrasts between natural and controlled mediator regimes. These results establish single-world representation and interpretation, not identification from observed data. The next subsection introduces the additional conditions required for observed-data identification.

\subsection{Identification of cross-world estimands via single-world assumptions} 

The previous subsections established that, under the necessity and sufficiency assumptions, cross-world quantities such as $\E[Y(a,M(a'))]$ can be represented as mixtures of single-world controlled counterfactual means within susceptibility-marker strata. Representation alone does not imply identification from observed data. Identification additionally requires conditions linking the controlled counterfactual means and the susceptibility-marker distributions to observed-data functionals. We make the following assumptions.

\begin{assumption}\label{ass:blinding}
    \textit{Marker blinding}: $S_M \perp A \mid X$.  
\end{assumption}

\begin{assumption}\label{ass:marker_exch}
    \textit{Marker-conditional single-world exchangeability}: For all $a,m$,
    \[
    \{Y(a,m), M(a)\} \perp A\mid X,S_M
    \quad\text{and}\quad
    Y(a,m)\perp M\mid A=a,X,S_M.
    \]
\end{assumption}

\begin{assumption}\label{ass:indep}
    \textit{Outcome-marker conditional independence}: 
    $Y \perp S_M \mid A,M,X$. 
\end{assumption} 

Assumption~\ref{ass:blinding} states that, conditional on $X$, treatment assignment does not select individuals according to their susceptibility marker. Assumption~\ref{ass:marker_exch} provides the single-world exchangeability conditions needed to identify controlled counterfactual means within susceptibility-marker strata. Assumption~\ref{ass:indep} states that, conditional on $(A,M,X)$, the observed outcome regression does not vary across marker strata. Consequently,
\[
\E[Y\mid A=a,M=m,X=x,S_M=s]
=
\E[Y\mid A=a,M=m,X=x].
\]
If $S_M$ were observed, Assumptions~\ref{ass:blinding} and~\ref{ass:indep} would impose empirically testable restrictions, whereas Assumption~\ref{ass:marker_exch} would remain a causal exchangeability condition. When $S_M$ is latent, all three conditions require substantive justification.

The two sufficiency conditions differ in which treatment arm reveals the susceptibility-marker strata. By Lemma~\ref{lem:sufficiency}, marker blinding, and consistency,
\begin{equation}
P(S_M=s)
=
\begin{cases}
\displaystyle
\int P(M=s\mid A=0,X=x)\,dP(x),
& \text{under sufficiency-under-control},\\[1em]
\displaystyle
\int P(M=s\mid A=1,X=x)\,dP(x),
& \text{under sufficiency-under-treatment},
\end{cases}
\label{eq:marker_prevalence}
\end{equation}
and
\begin{equation}
dP(x\mid S_M=s)
=
\begin{cases}
\displaystyle
\frac{
P(M=s\mid A=0,X=x)\,dP(x)
}{
\int P(M=s\mid A=0,X=x')\,dP(x')
},
& \text{under sufficiency-under-control},\\[1.5em]
\displaystyle
\frac{
P(M=s\mid A=1,X=x)\,dP(x)
}{
\int P(M=s\mid A=1,X=x')\,dP(x')
},
& \text{under sufficiency-under-treatment}.
\end{cases}
\label{eq:marker_covariate_distributions}
\end{equation}
Thus, although the formulas below retain $P(X\mid S_M=s)$ to make their susceptibility-stratum interpretation explicit, this distribution is identified from the mediator distribution in the corresponding treatment arm.

\begin{theorem}
\label{thm:id_single_world}
Suppose Assumptions~\ref{ass:consistency}--\ref{ass:single_world_ignor}, \ref{ass:necessity}, and \ref{ass:blinding}--\ref{ass:indep} hold.
\\[0.2cm]
Under Assumption~\ref{ass:suff_control},
\begin{equation}
\E[Y(a,M(0))]
=
\int\sum_{m=0}^1
\E[Y\mid A=a,M=m,X=x]\,
P(M=m\mid A=0,X=x)\,dP(x).
\label{eq:id_nested_control}
\end{equation}
Provided $P(S_M=s)>0$, the conditional controlled direct effects are identified, for $s\in\{0,1\}$, by
\begin{equation}
\Delta^\text{dir-c}_s
=
\int
\left\{
\E[Y\mid A=1,M=s,X=x]
-
\E[Y\mid A=0,M=s,X=x]
\right\}
dP(x\mid S_M=s),
\label{eq:id_dir_c}
\end{equation}
where the first line of \eqref{eq:marker_covariate_distributions} identifies the covariate distribution within each marker stratum. Provided $P(S_M=1)>0$, the susceptibility-stratum-specific mediator-regime contrast is identified by
\begin{align}
\Delta^\text{indir-c}_1
&=
\left\{
\int P(M=1\mid A=0,x)\,dP(x)
\right\}^{-1}
\notag\\[-0.2em]
&\quad\times
\int\sum_{m=0}^1
\E[Y\mid A=1,m,x]
\, 
\left\{
P(m\mid A=1,x)
-
P(m\mid A=0,x)
\right\}
dP(x).
\label{eq:id_indir_c}
\end{align}
Consequently, the pure natural direct effect and total natural indirect effect defined in \eqref{eq:natural_effect_decompositions} are identified, respectively, by
$\Delta^\text{dir-c}_1P(S_M=1)
+
\Delta^\text{dir-c}_0P(S_M=0),
$ and $\Delta^\text{indir-c}P(S_M=1).$
\\[0.2cm]
Under Assumption~\ref{ass:suff_trt},
\begin{equation}
\E[Y(a,M(1))]
=
\int\sum_{m=0}^1
\E[Y\mid A=a,M=m,X=x]\,
P(M=m\mid A=1,X=x)\,dP(x).
\label{eq:id_nested_treatment}
\end{equation}
Provided $P(S_M=s)>0$, the conditional controlled direct effects are identified, for $s\in\{0,1\}$, by
\begin{equation}
\Delta^\text{dir-t}_s
=
\int
\left\{
\E[Y\mid A=1,M=s,X=x]
-
\E[Y\mid A=0,M=s,X=x]
\right\}
dP(x\mid S_M=s),
\label{eq:id_dir_t}
\end{equation}
where the second line of \eqref{eq:marker_covariate_distributions} identifies the covariate distribution within each marker stratum. Provided $P(S_M=1)>0$, the susceptibility-stratum-specific mediator-regime contrast is identified by
\begin{align}
\Delta^\text{indir-t}_1
&=
\left\{
\int P(M=1\mid A=1,x)\,dP(x)
\right\}^{-1}
\notag\\[-0.2em]
&\quad\times
\int\sum_{m=0}^1
\E[Y\mid A=0,m,x]
\, 
\left\{
P(m\mid A=1,x)
-
P(m\mid A=0,x)
\right\}
dP(x).
\label{eq:id_indir_t}
\end{align}
Consequently, the total natural direct effect and pure natural indirect effect are identified, respectively, by
$
\Delta^\text{dir-t}_1P(S_M=1)
+
\Delta^\text{dir-t}_0P(S_M=0),
$ and 
$\Delta^\text{indir-t}P(S_M=1).
$
\end{theorem}

See Appendix~\ref{app:proofs_id_single_world} for a proof.

Theorem~\ref{thm:id_single_world} completes the second step of the argument. The preceding subsections represented and interpreted the natural effects as mixtures of single-world causal contrasts within susceptibility-marker strata. The theorem shows that, under the additional assumptions above, the controlled counterfactual means, marker-stratum distributions, and resulting natural effects are functionals of the observed-data distribution. Although $S_M$ need not be observed, its prevalence and its stratum-specific covariate distributions are recovered from the mediator distribution in the treatment arm under which Lemma~\ref{lem:sufficiency} equates the marker with the mediator.

\section{Implications for principal causal effects}
\label{sec:principal_effects}

The preceding sections used the susceptibility marker to represent nested mediation counterfactuals as mixtures of single-world quantities. The same construction has a further implication for principal stratification. In classical principal stratification, principal strata are defined by the joint potential outcomes $(M(1),M(0))$ \citep{frangakis2002principal,vanderweele2011principal}. Thus, an observed marker that equals either $M(0)$ or $M(1)$ reveals one otherwise latent coordinate of the principal stratum. The purpose of this section is to determine how much of the principal-stratification problem this information resolves and which components remain unidentified. The marker does not, by itself, identify every principal causal effect; rather, it reduces the latent structure of the problem and isolates the additional assumption needed for full identification.

For a binary mediator, there are four such strata: $(1,1)$, $(1,0)$, $(0,1)$, and $(0,0)$. Because both $M(1)$ and $M(0)$ are ordinarily not observed for the same individual, principal-stratum membership is latent and defined across treatment worlds. Consequently, estimands of the form
\[
\E[Y(1)-Y(0)\mid M(1)=m_1,M(0)=m_0]
\]
are generally not identified without additional assumptions, such as principal ignorability, monotonicity, exclusion restrictions, and/or parametric restrictions.

An important example arises in studies of vaccine effects on post-infection outcomes. In this setting, $A=1$ denotes vaccination and $M=1$ denotes infection. \citet{codi2026causal} study causal vaccine effects among individuals who would become infected in the absence of vaccination, defining this group, characterized by $M(0)=1$, as the \emph{Naturally Infected}. Under the prevention monotonicity $M(1)\leq M(0)$, this population contains both individuals who would become infected regardless of vaccination, with $(M(1),M(0))=(1,1)$, and individuals whose infection would be prevented by vaccination, with $(M(1),M(0))=(0,1)$. The former constitute the commonly studied \emph{Doomed} stratum. \citet{codi2026causal} derive bounds and point-identification results for causal effects in the broader Naturally Infected population under exclusion restrictions and partial principal ignorability conditions. They also provide a single-world interpretation based on exposure to a sufficiently infectious dose.

The vaccine construction of \citet{codi2026causal} can be expressed within the susceptibility-marker framework. To make the connection explicit, suppose that an observed marker $S_M$ encodes an exposure--susceptibility condition of the type considered by \citet{codi2026causal} and satisfies necessity and sufficiency-under-control. Lemma~\ref{lem:sufficiency} then gives $S_M=M(0)$, so the Naturally Infected population defined by \citet{codi2026causal} corresponds to $S_M=1$. Among vaccinated individuals in this population, observed infection status further distinguishes the two relevant principal strata: $(A,M,S_M)=(1,1,1)$ identifies the always-infected stratum $(1,1)$, whereas $(A,M,S_M)=(1,0,1)$ identifies the stratum $(0,1)$ whose infection is prevented by vaccination.

This mapping illustrates how an observed susceptibility marker permits partial reconstruction of $(M(1),M(0))$ from the single-world quantities $(A,M,S_M)$. To characterize what this partial reconstruction contributes to identification, we separate the problem into identifying the principal-stratum probabilities and identifying the potential-outcome means within each stratum.

Principal causal effects are contrasts of principal-stratum-specific potential-outcome means. For a stratum $(m_1,m_0)$ with positive probability,
\begin{equation}
\begin{aligned}
&\E[Y(a)\mid M(1)=m_1,M(0)=m_0] \\
&\quad=
\frac{
\displaystyle
\int
\E[Y(a)\mid M(1)=m_1,M(0)=m_0,X=x]\,
P(M(1)=m_1,M(0)=m_0\mid X=x)\,dP(x)
}{
\displaystyle
\int
P(M(1)=m_1,M(0)=m_0\mid X=x)\,dP(x)
}.
\end{aligned}
\label{eq:principal_mean_marginal}
\end{equation}
Thus, identification of a principal causal effect requires identification of both the principal-stratum probabilities
\[
P(M(1)=m_1,M(0)=m_0\mid X=x)
\]
and the conditional principal-stratum means
\[
\E[Y(a)\mid M(1)=m_1,M(0)=m_0,X=x].
\]

The two sufficiency conditions imply opposite monotonicity directions. Necessity and sufficiency-under-treatment imply $M(0)\leq M(1)$ and therefore restrict the feasible values of $(M(1),M(0))$ to
\[
(0,0),\quad (1,0),\quad\text{and}\quad (1,1).
\]
Here, $(1,0)$ is the treatment-induced stratum. Conversely, necessity and sufficiency-under-control imply $M(1)\leq M(0)$ and restrict the feasible strata to
\[
(0,0),\quad (0,1),\quad\text{and}\quad (1,1),
\]
corresponding, respectively, to the never-$1$, treatment-prevented, and always-$1$ strata. Analogous arguments apply under the opposite monotonicity direction. We focus on $M(1)\leq M(0)$ because it corresponds directly to the vaccine-infection setting above.

For the identification results below, we retain Assumptions~\ref{ass:consistency}--\ref{ass:single_world_ignor}, \ref{ass:necessity}, \ref{ass:suff_control}, and \ref{ass:blinding}--\ref{ass:indep}, and suppose that $S_M$ is observed. Define
\[
p_{ma}(x)=P(M=m\mid A=a,X=x),
\qquad
\mu_{ma}(x)=\E[Y\mid M=m,A=a,X=x],
\]
and
\[
\mu_{m,a,s}(x)
=
\E[Y\mid M=m,A=a,S_M=s,X=x].
\]
Assumption~\ref{ass:indep} implies $\mu_{m,a,s}(x)=\mu_{ma}(x)$ whenever the corresponding conditioning event has positive probability.

Under $M(1)\leq M(0)$, the conditional principal-stratum probabilities are
\begin{equation}
\begin{aligned}
P(M(1)=1,M(0)=1\mid X=x)
&=p_{11}(x),\\
P(M(1)=0,M(0)=0\mid X=x)
&=p_{00}(x),\\
P(M(1)=0,M(0)=1\mid X=x)
&=p_{10}(x)-p_{11}(x).
\end{aligned}
\label{eq:principal_stratum_probabilities}
\end{equation}
In particular, the monotonicity restriction implies the observable inequality $p_{11}(x)\leq p_{10}(x)$.

Two principal-stratum means are identified directly from the treatment arms in which the corresponding mediator value reveals the stratum:
\begin{equation}
\begin{aligned}
\E[Y(1)\mid M(1)=1,M(0)=1,X=x]
&=\mu_{11}(x),\\
\E[Y(0)\mid M(1)=0,M(0)=0,X=x]
&=\mu_{00}(x).
\end{aligned}
\label{eq:principal_means_direct}
\end{equation}
The observed marker additionally separates the two strata with $M(1)=0$. Specifically,
\begin{equation}
\begin{aligned}
\E[Y(1)\mid M(1)=0,M(0)=0,X=x]
&=
\mu_{0,1,0}(x)
=
\mu_{01}(x),\\
\E[Y(1)\mid M(1)=0,M(0)=1,X=x]
&=
\mu_{0,1,1}(x)
=
\mu_{01}(x).
\end{aligned}
\label{eq:principal_means_marker}
\end{equation}
The first equality in each line follows from the observed-marker mapping, consistency, and marker-conditional exchangeability. The second follows from outcome-marker conditional independence. Without Assumption~\ref{ass:indep}, the two means would remain identified by the corresponding $S_M$-specific outcome regressions but would not generally be equal.

Table~\ref{tab:mapping} summarizes how the observed values of $(A,M,S_M)$ map to the feasible principal strata.

\begin{table}[t]
\centering
\caption{Mapping between observed $(A,M,S_M)$ and feasible principal strata $(M(1),M(0))$ under $S_M=M(0)$ and $M(1)\leq M(0)$.}
\label{tab:mapping}
\begin{tabular}{ccccl}
\toprule
$A$ & $M$ & $S_M$ & $(M(1),M(0))$ & Interpretation \\
\midrule
1 & 1 & 1 & $(1,1)$ & Always-$1$ \\
1 & 0 & 1 & $(0,1)$ & Treatment-prevented \\
1 & 0 & 0 & $(0,0)$ & Never-$1$ \\
0 & 0 & 0 & $(0,0)$ & Never-$1$ \\
0 & 1 & 1 & $(0,1)$ or $(1,1)$ & Treatment-prevented or always-$1$ \\
\bottomrule
\end{tabular}
\end{table}

The remaining unidentified means are
\begin{equation}
\begin{aligned}
&\E[Y(0)\mid M(1)=1,M(0)=1,X=x],\\
&\E[Y(0)\mid M(1)=0,M(0)=1,X=x].
\end{aligned}
\label{eq:principal_remaining_means}
\end{equation}
These two quantities are related through the observed outcome regression among untreated individuals with $M=1$. Provided
\[
p_{10}(x)-p_{11}(x)>0,
\]
we have
\begin{equation}
\begin{aligned}
&\E[Y(0)\mid M(1)=0,M(0)=1,X=x]\\
&\quad=
\frac{
p_{10}(x)\mu_{10}(x)
-
p_{11}(x)
\E[Y(0)\mid M(1)=1,M(0)=1,X=x]
}{
p_{10}(x)-p_{11}(x)
}.
\end{aligned}
\label{eq:principal_prevented_mean}
\end{equation}
Thus, four of the six conditional principal-stratum means are identified directly. The two remaining means in \eqref{eq:principal_remaining_means} are linked through \eqref{eq:principal_prevented_mean}, so the unidentified portion of the problem is one-dimensional. It may be indexed by
\[
\E[Y(0)\mid M(1)=1,M(0)=1,X=x].
\]
Once this mean is identified, \eqref{eq:principal_prevented_mean} identifies the untreated outcome mean in the treatment-prevented stratum, and all principal causal effects follow.

For comparison, principal ignorability,
\[
Y(a)\perp(M(1),M(0))\mid X,
\]
would set every conditional principal-stratum mean under treatment level $a$ equal to $\E[Y(a)\mid X=x]$, which is identified by $\E[Y\mid A=a,X=x]$ under consistency and treatment exchangeability. The observed-marker construction requires neither this full equality across principal strata nor the corresponding principal-ignorability assumption. Instead, it reduces the unidentified part of the problem to one conditional mean.

One possible way to identify this remaining mean is a context-specific exclusion restriction.

\begin{assumption}\label{ass:ex_restriction}
    \textit{Always-$1$ conditional-mean exclusion restriction}: For all $x$,
    \[
    \E[Y(0,1)\mid M(1)=1,M(0)=1,X=x]
    =
    \E[Y(1,1)\mid M(1)=1,M(0)=1,X=x].
    \]
\end{assumption}

Assumption~\ref{ass:ex_restriction} states that, within the always-$1$ stratum, treatment has no conditional average effect on the outcome when the mediator is fixed at $1$. It is weaker than imposing $Y(0,1)=Y(1,1)$ for every individual. Under this restriction,
\begin{equation}
\E[Y(0)\mid M(1)=1,M(0)=1,X=x]
=
\mu_{11}(x).
\label{eq:principal_always_mean}
\end{equation}
Substituting \eqref{eq:principal_always_mean} into \eqref{eq:principal_prevented_mean} identifies the remaining mean:
\begin{equation}
\begin{aligned}
&\E[Y(0)\mid M(1)=0,M(0)=1,X=x]\\
&\quad=
\frac{
p_{10}(x)\mu_{10}(x)-p_{11}(x)\mu_{11}(x)
}{
p_{10}(x)-p_{11}(x)
}.
\end{aligned}
\label{eq:principal_prevented_mean_identified}
\end{equation}

To state the resulting principal causal effects explicitly, let
\[
\tau_{m_1m_0}(x)
=
\E[Y(1)-Y(0)\mid M(1)=m_1,M(0)=m_0,X=x].
\]
Under Assumption~\ref{ass:ex_restriction},
\begin{equation}
\begin{aligned}
\tau_{11}(x)
&=0,\\
\tau_{00}(x)
&=\mu_{01}(x)-\mu_{00}(x),\\
\tau_{01}(x)
&=
\mu_{01}(x)
-
\frac{
p_{10}(x)\mu_{10}(x)-p_{11}(x)\mu_{11}(x)
}{
p_{10}(x)-p_{11}(x)
}.
\end{aligned}
\label{eq:identified_principal_effects}
\end{equation}
The corresponding marginal principal causal effects follow by averaging each conditional effect over the covariate distribution within its principal stratum, as in \eqref{eq:principal_mean_marginal}.

The principal-stratification analysis therefore shows that the susceptibility-marker construction has implications beyond mediation. In the mediation results, the equality between $S_M$ and the mediator under a reference treatment allowed nested counterfactual means to be represented as mixtures of single-world quantities. Here, the same equality reveals one coordinate of the principal stratum. Under sufficiency-under-control, observing $S_M=M(0)$, together with the observed mediator among treated individuals, identifies all feasible principal-stratum probabilities and four of the six principal-stratum-specific outcome means. The two remaining means are linked through an observed-data mixture, leaving one unidentified conditional mean that may be addressed through Assumption~\ref{ass:ex_restriction}. The vaccine setting of \citet{codi2026causal} provides a concrete instance of this implication: their Naturally Infected population corresponds to $S_M=1$, while infection status among vaccine recipients separates the always-infected and vaccine-prevented strata. This interpretation is due to \citet{codi2026causal}. Here, it illustrates the broader point that a susceptibility marker does not eliminate the need for additional assumptions, but can reduce and localize the unidentified part of a principal-stratification problem.

See Appendix~\ref{app:proofs_principalstrata} for details.

\section{Conclusions}
\label{sec:conc}

We have shown how cross-world mediation counterfactuals can be represented as mixtures of single-world quantities defined with respect to susceptibility markers. Under necessity and sufficiency-under-control, the marker equals $M(0)$, yielding a single-world mixture representation of the decomposition into the pure natural direct effect and the total natural indirect effect. Under necessity and sufficiency-under-treatment, the marker equals $M(1)$, yielding the corresponding representation of the decomposition into the total natural direct effect and the pure natural indirect effect. In both cases, the direct components are conditional controlled direct effects, whereas the mediated components are susceptibility-stratum-specific contrasts between natural and controlled mediator regimes. These restrictions do not redefine the natural effects; they provide single-world representations and interpretations of the original cross-world estimands. Under the additional marker-blinding, marker-conditional exchangeability, and outcome-marker conditional-independence assumptions, together with consistency, positivity, and single-world ignorability, the resulting mixtures are identified from observed data.

The same construction also has implications for principal causal effects. When the susceptibility marker is observed, its equality with one mediator potential outcome reveals one coordinate of the principal stratum. In the sufficiency-under-control case examined here, this identifies all principal-stratum probabilities and four of the six principal-stratum-specific outcome means without invoking principal ignorability. The remaining two means are linked through an observed-data mixture, leaving one unidentified conditional mean that might be identified through a more targeted, context-specific restriction, such as the exclusion restriction considered in Section~\ref{sec:principal_effects}. Thus, the marker does not remove the cross-world definition of principal strata, but it reduces their latent structure and localizes the additional assumption required for full identification.

The susceptibility-marker approach does not eliminate the need for scientific judgment. Necessity and sufficiency impose deterministic relations between the marker and the mediator, while the identification conditions impose additional restrictions on the relationships among treatment, mediator, marker, and outcome. Each must be justified in the scientific context. The examples considered here illustrate settings in which biological, behavioral, or environmental mechanisms may support these conditions. When they are credible, susceptibility markers clarify which cross-world quantities admit single-world representations, how their components should be interpreted, and what additional assumptions are required for observed-data identification.

Although the main development focuses on a single mediator, the construction is not limited to this setting. In the Supplementary Material, we extend it to ordered mediators and apply it to mediation settings with exposure-induced mediator--outcome confounding. Together, these results show that susceptibility markers provide a coherent strategy for representing part of the latent cross-world structure underlying classical mediation estimands through scientifically interpretable single-world variables, while preserving the original estimands and making the remaining identifying assumptions explicit.

\section*{Acknowledgments}

We thank Mats Stensrud for helpful discussions during the early development of this work. R.N. also thanks the Isaac Newton Institute for Mathematical Sciences, Cambridge, for support and hospitality during the programme \textit{Causal inference: From theory to practice and back again}, and the participants in talks, panel discussions, and follow-up roundtables for stimulating discussions on mediation that contributed to this work.

\vspace{1cm}

\bibliographystyle{plainnat}  
\bibliography{references} 

\pagebreak 
\appendix 

\renewcommand{\thesection}{S\arabic{section}}
\renewcommand{\theequation}{S\arabic{equation}}
\setcounter{section}{0}
\setcounter{equation}{0}

\noindent {\LARGE \bf Supplementary Materials}
\vspace{0.75cm}

The supplementary material is organized as follows. Appendix~\ref{app:proofs} provides proofs and supporting derivations for the single-mediator results presented in the main text, including the results concerning principal stratification. Appendix~\ref{app:multiple_mediators} extends the susceptibility-marker formulation to ordered mediators and establishes a single-world representation and identification result for path-specific effects. Appendix~\ref{app:exposure_induced_confounding} examines the implications of this construction in the presence of exposure-induced mediator--outcome confounding and clarifies its connection to monotonicity-based identification. 

\vspace{1.5cm}
\startcontents[supp]
{ 
\renewcommand{\baselinestretch}{1.2}\selectfont
\printcontents[supp]{}{1}{}
}

\newpage
\section{Proofs for the single-mediator results}
\label{app:proofs}

\subsection{Proof of Lemma~\ref{lem:sufficiency}}
\label{app:proofs_sufficiency}

Under necessity and sufficiency-under-control, for almost every $x$,
\begin{align*}
P(M(0)=S_M\mid X=x)
={}&
P(M(0)=0\mid S_M=0,X=x) \, P(S_M=0\mid X=x)\\
&+
P(M(0)=1\mid S_M=1,X=x) \, P(S_M=1\mid X=x).
\end{align*}
Necessity implies $P(M(0)=0\mid S_M=0,X=x)=1$, while sufficiency-under-control implies $P(M(0)=1\mid S_M=1,X=x)=1.$ Therefore, $P(M(0)=S_M\mid X=x)=1,$ and hence $M(0)=S_M$ almost surely.

Similarly, under necessity and sufficiency-under-treatment,
\begin{align*}
P(M(1)=S_M\mid X=x)
={}&
P(M(1)=0\mid S_M=0,X=x) \, P(S_M=0\mid X=x)\\
&+
P(M(1)=1\mid S_M=1,X=x) \, P(S_M=1\mid X=x).
\end{align*}
Necessity implies $P(M(1)=0\mid S_M=0,X=x)=1,$ while sufficiency-under-treatment implies $P(M(1)=1\mid S_M=1,X=x)=1.$ Therefore, $P(M(1)=S_M\mid X=x)=1,$ and hence $M(1)=S_M$ almost surely.

\hfill\(\square\)





\subsection{Proof of Theorem~\ref{thm:id_single_world}}
\label{app:proofs_id_single_world}

We first establish an identification result used under both sufficiency conditions. For all $a,m,s$, Assumptions~\ref{ass:consistency}, \ref{ass:marker_exch}, and~\ref{ass:indep} imply
\begin{align}
\E[Y(a,m)\mid S_M=s,X=x]
&=
\E[Y(a,m)\mid A=a,M=m,S_M=s,X=x]
\notag\\
&=
\E[Y\mid A=a,M=m,S_M=s,X=x]
\notag\\
&=
\E[Y\mid A=a,M=m,X=x].
\label{eq:proof_controlled_mean}
\end{align}
The first equality follows by applying the two marker-conditional exchangeability conditions in Assumption~\ref{ass:marker_exch}: the first permits conditioning on $A=a$, and the second permits further conditioning on $M=m$. The second equality follows from consistency in Assumption~\ref{ass:consistency}, and the third follows from outcome-marker conditional independence in Assumption~\ref{ass:indep}. Thus, a controlled counterfactual mean within a marker stratum is identified by an observed outcome regression that does not condition on the possibly latent marker.

\subsubsection{Sufficiency-under-control}
\label{app:proofs_id_single_world_suff_under_control}

Under necessity and sufficiency-under-control, Lemma~\ref{lem:sufficiency} gives $S_M=M(0)$. Marker blinding and consistency therefore imply
\begin{align}
P(S_M=s\mid X=x)
&=
P(S_M=s\mid A=0,X=x)
\notag\\
&=
P(M=s\mid A=0,X=x).
\label{eq:proof_marker_control}
\end{align}
The first equality follows from Assumption~\ref{ass:blinding}. For the second equality, $S_M=M(0)$ and, among individuals with $A=0$, consistency gives $M=M(0)=S_M$. Integrating \eqref{eq:proof_marker_control} over $P(x)$ gives
\[
P(S_M=s)
=
\int P(M=s\mid A=0,X=x)\,dP(x),
\]
which is the first line of \eqref{eq:marker_prevalence}. Bayes' rule then gives
\[
dP(x\mid S_M=s)
=
\frac{
P(M=s\mid A=0,X=x)\,dP(x)
}{
\int P(M=s\mid A=0,X=x')\,dP(x')
},
\]
which is the first line of \eqref{eq:marker_covariate_distributions}.

For any $a\in\{0,1\}$, the nested counterfactual mean can be written as
\begin{align}
\E[Y(a,M(0))]
&=
\int\sum_{m=0}^1
\E[Y(a,m)\mid M(0)=m,X=x]\,
P(M(0)=m\mid X=x)\,dP(x)
\notag\\
&=
\int\sum_{m=0}^1
\E[Y(a,m)\mid S_M=m,X=x]\,
P(S_M=m\mid X=x)\,dP(x)
\notag\\
&=
\int\sum_{m=0}^1
\E[Y(a,m)\mid S_M=m,X=x]\,
P(M=m\mid A=0,X=x)\,dP(x)
\notag\\
&=
\int\sum_{m=0}^1
\E[Y\mid A=a,M=m,X=x]\,
P(M=m\mid A=0,X=x)\,dP(x).
\label{eq:proof_nested_control}
\end{align}
The first equality follows from the law of total expectation and composition. The second uses $S_M=M(0)$, the third follows from \eqref{eq:proof_marker_control}, and the final equality follows from \eqref{eq:proof_controlled_mean}. This proves \eqref{eq:id_nested_control}.

For $s\in\{0,1\}$, the conditional controlled direct effect is
\begin{align}
\Delta^\text{dir-c}_s
&=
\E[Y(1,s)-Y(0,s)\mid S_M=s]
\notag\\
&=
\int
\left\{
\E[Y(1,s)\mid S_M=s,X=x]
-
\E[Y(0,s)\mid S_M=s,X=x]
\right\}
dP(x\mid S_M=s)
\notag\\
&=
\int
\left\{
\E[Y\mid A=1,M=s,X=x]
-
\E[Y\mid A=0,M=s,X=x]
\right\}
dP(x\mid S_M=s).
\label{eq:proof_dir_control}
\end{align}
The second equality follows from the law of total expectation, and the final equality follows from \eqref{eq:proof_controlled_mean}. Because $P(X\mid S_M=s)$ is identified by the first line of \eqref{eq:marker_covariate_distributions}, this proves \eqref{eq:id_dir_c} for both $\Delta^\text{dir-c}_1$ and $\Delta^\text{dir-c}_0$.

It remains to identify $\Delta^\text{indir-c}_1$. By composition,
\[
Y(1,M(1))=Y(1),
\]
so the total natural indirect effect can be written as
\begin{align*}
\E[Y(1,M(1))-Y(1,M(0))]
&=
\E[Y(1)]-\E[Y(1,M(0))].
\end{align*}
Under consistency and single-world ignorability,
\begin{align}
\E[Y(1)]
&=
\int \E[Y\mid A=1,X=x]\,dP(x)
\notag\\
&=
\int\sum_{m=0}^1
\E[Y\mid A=1,M=m,X=x]\,
P(M=m\mid A=1,X=x)\,dP(x).
\label{eq:proof_y1}
\end{align}
Setting $a=1$ in \eqref{eq:proof_nested_control} gives
\begin{align}
\E[Y(1,M(0))]
&=
\int\sum_{m=0}^1
\E[Y\mid A=1,M=m,X=x]\,
P(M=m\mid A=0,X=x)\,dP(x).
\label{eq:proof_y1m0}
\end{align}
Subtracting \eqref{eq:proof_y1m0} from \eqref{eq:proof_y1} yields
\begin{align}
&\E[Y(1,M(1))-Y(1,M(0))]
\notag\\
&\quad=
\int\sum_{m=0}^1
\E[Y\mid A=1,M=m,X=x]
\notag\\[-0.2em]
&\qquad\quad\times
\left\{
P(M=m\mid A=1,X=x)
-
P(M=m\mid A=0,X=x)
\right\}
dP(x).
\label{eq:proof_tnie}
\end{align}

The single-world representation in \eqref{eq:indir_eff_suff_under_control} gives
\[
\E[Y(1,M(1))-Y(1,M(0))]
=
\Delta^\text{indir-c}_1P(S_M=1).
\]
Moreover, by the first line of \eqref{eq:marker_prevalence},
\[
P(S_M=1)
=
\int P(M=1\mid A=0,X=x)\,dP(x).
\]
Provided $P(S_M=1)>0$, dividing \eqref{eq:proof_tnie} by this marker-stratum probability gives
\begin{align*}
\Delta^\text{indir-c}_1
&=
\left\{
\int P(M=1\mid A=0,X=x)\,dP(x)
\right\}^{-1}
\\[-0.2em]
&\quad\times
\int\sum_{m=0}^1
\E[Y\mid A=1,M=m,X=x]
\\[-0.2em]
&\qquad\quad\times
\left\{
P(M=m\mid A=1,X=x)
-
P(M=m\mid A=0,X=x)
\right\}
dP(x),
\end{align*}
which proves \eqref{eq:id_indir_c}.

Finally, the single-world representations in
\eqref{eq:controlled_dir_eff_suff_under_control} and
\eqref{eq:indir_eff_suff_under_control} give
\begin{align*}
\E[Y(1,M(0))-Y(0,M(0))]
&=
\Delta^\text{dir-c}_1P(S_M=1)
+
\Delta^\text{dir-c}_0P(S_M=0),\\
\E[Y(1,M(1))-Y(1,M(0))]
&=
\Delta^\text{indir-c}P(S_M=1).
\end{align*}
Because every quantity on the right-hand sides has been identified above, this concludes the proof.

\subsubsection{Sufficiency-under-treatment}
\label{app:proofs_id_single_world_suff_under_trt}

Under necessity and sufficiency-under-treatment, Lemma~\ref{lem:sufficiency} gives $S_M=M(1)$. Marker blinding and consistency therefore imply
\begin{align}
P(S_M=s\mid X=x)
&=
P(S_M=s\mid A=1,X=x)
\notag\\
&=
P(M=s\mid A=1,X=x).
\label{eq:proof_marker_treatment}
\end{align}
The first equality follows from Assumption~\ref{ass:blinding}. For the second equality, $S_M=M(1)$ and, among individuals with $A=1$, consistency gives $M=M(1)=S_M$. Integrating \eqref{eq:proof_marker_treatment} over $P(x)$ gives
\[
P(S_M=s)
=
\int P(M=s\mid A=1,X=x)\,dP(x),
\]
which is the second line of \eqref{eq:marker_prevalence}. Bayes' rule then gives
\[
dP(x\mid S_M=s)
=
\frac{
P(M=s\mid A=1,X=x)\,dP(x)
}{
\int P(M=s\mid A=1,X=x')\,dP(x')
},
\]
which is the second line of \eqref{eq:marker_covariate_distributions}.

For any $a\in\{0,1\}$, the nested counterfactual mean can be written as
\begin{align}
\E[Y(a,M(1))]
&=
\int\sum_{m=0}^1
\E[Y(a,m)\mid M(1)=m,X=x]\,
P(M(1)=m\mid X=x)\,dP(x)
\notag\\
&=
\int\sum_{m=0}^1
\E[Y(a,m)\mid S_M=m,X=x]\,
P(S_M=m\mid X=x)\,dP(x)
\notag\\
&=
\int\sum_{m=0}^1
\E[Y(a,m)\mid S_M=m,X=x]\,
P(M=m\mid A=1,X=x)\,dP(x)
\notag\\
&=
\int\sum_{m=0}^1
\E[Y\mid A=a,M=m,X=x]\,
P(M=m\mid A=1,X=x)\,dP(x).
\label{eq:proof_nested_treatment}
\end{align}
The first equality follows from the law of total expectation and composition. The second uses $S_M=M(1)$, the third follows from \eqref{eq:proof_marker_treatment}, and the final equality follows from \eqref{eq:proof_controlled_mean}. This proves \eqref{eq:id_nested_treatment}.

For $s\in\{0,1\}$, the conditional controlled direct effect is
\begin{align}
\Delta^\text{dir-t}_s
&=
\E[Y(1,s)-Y(0,s)\mid S_M=s]
\notag\\
&=
\int
\left\{
\E[Y(1,s)\mid S_M=s,X=x]
-
\E[Y(0,s)\mid S_M=s,X=x]
\right\}
dP(x\mid S_M=s)
\notag\\
&=
\int
\left\{
\E[Y\mid A=1,M=s,X=x]
-
\E[Y\mid A=0,M=s,X=x]
\right\}
dP(x\mid S_M=s).
\label{eq:proof_dir_treatment}
\end{align}
The second equality follows from the law of total expectation, and the final equality follows from \eqref{eq:proof_controlled_mean}. Because $P(X\mid S_M=s)$ is identified by the second line of \eqref{eq:marker_covariate_distributions}, this proves \eqref{eq:id_dir_t} for both $\Delta^\text{dir-t}_1$ and $\Delta^\text{dir-t}_0$.

It remains to identify $\Delta^\text{indir-t}_1$. By composition,
\[
Y(0,M(0))=Y(0),
\]
so the pure natural indirect effect can be written as
\begin{align*}
\E[Y(0,M(1))-Y(0,M(0))]
&=
\E[Y(0,M(1))]-\E[Y(0)].
\end{align*}
Setting $a=0$ in \eqref{eq:proof_nested_treatment} gives
\begin{align}
\E[Y(0,M(1))]
&=
\int\sum_{m=0}^1
\E[Y\mid A=0,M=m,X=x]\,
P(M=m\mid A=1,X=x)\,dP(x).
\label{eq:proof_y0m1}
\end{align}
Under consistency and single-world ignorability,
\begin{align}
\E[Y(0)]
&=
\int \E[Y\mid A=0,X=x]\,dP(x)
\notag\\
&=
\int\sum_{m=0}^1
\E[Y\mid A=0,M=m,X=x]\,
P(M=m\mid A=0,X=x)\,dP(x).
\label{eq:proof_y0}
\end{align}
Subtracting \eqref{eq:proof_y0} from \eqref{eq:proof_y0m1} yields
\begin{align}
&\E[Y(0,M(1))-Y(0,M(0))]
\notag\\
&\quad=
\int\sum_{m=0}^1
\E[Y\mid A=0,M=m,X=x]
\notag\\[-0.2em]
&\qquad\quad\times
\left\{
P(M=m\mid A=1,X=x)
-
P(M=m\mid A=0,X=x)
\right\}
dP(x).
\label{eq:proof_pnie}
\end{align}

The single-world representation in \eqref{eq:indir_eff_suff_under_trt} gives
\[
\E[Y(0,M(1))-Y(0,M(0))]
=
\Delta^\text{indir-t}_1P(S_M=1).
\]
Moreover, by the second line of \eqref{eq:marker_prevalence},
\[
P(S_M=1)
=
\int P(M=1\mid A=1,X=x)\,dP(x).
\]
Provided $P(S_M=1)>0$, dividing \eqref{eq:proof_pnie} by this marker-stratum probability gives
\begin{align*}
\Delta^\text{indir-t}_1
&=
\left\{
\int P(M=1\mid A=1,X=x)\,dP(x)
\right\}^{-1}
\\[-0.2em]
&\quad\times
\int\sum_{m=0}^1
\E[Y\mid A=0,M=m,X=x]
\\[-0.2em]
&\qquad\quad\times
\left\{
P(M=m\mid A=1,X=x)
-
P(M=m\mid A=0,X=x)
\right\}
dP(x),
\end{align*}
which proves \eqref{eq:id_indir_t}.

Finally, the single-world representations in
\eqref{eq:dir_eff_suff_under_trt} and
\eqref{eq:indir_eff_suff_under_trt} give
\begin{align*}
\E[Y(1,M(1))-Y(0,M(1))]
&=
\Delta^\text{dir-t}_1P(S_M=1)
+
\Delta^\text{dir-t}_0P(S_M=0),\\
\E[Y(0,M(1))-Y(0,M(0))]
&=
\Delta^\text{indir-t}P(S_M=1).
\end{align*}
Because every quantity on the right-hand sides has been identified above, this concludes the proof. 

\hfill\(\square\)

\subsection{Proof of identification of principal effects}
\label{app:proofs_principalstrata}

We work under necessity and sufficiency-under-control, so Lemma~\ref{lem:sufficiency} gives $S_M=M(0)$ and
\[
M(1)\leq M(0).
\]
We also retain the consistency, positivity, exchangeability, and outcome-marker conditional-independence conditions stated in the main text.

Under $M(1)\leq M(0)$, the event $\{M(1)=1\}$ implies $\{M(0)=1\}$. Therefore,
\begin{align*}
P(M(1)=1,M(0)=1\mid X=x)
&=
P(M(1)=1\mid X=x)\\
&=
P(M=1\mid A=1,X=x)\\
&=
p_{11}(x),
\end{align*}
where the second equality follows from consistency and treatment exchangeability. Similarly, the event $\{M(0)=0\}$ implies $\{M(1)=0\}$, so
\begin{align*}
P(M(1)=0,M(0)=0\mid X=x)
&=
P(M(0)=0\mid X=x)\\
&=
P(M=0\mid A=0,X=x)\\
&=
p_{00}(x).
\end{align*}
The remaining principal-stratum probability is
\begin{align*}
P(M(1)=0,M(0)=1\mid X=x)
&=
P(M(0)=1\mid X=x)
-
P(M(1)=1\mid X=x)\\
&=
p_{10}(x)-p_{11}(x).
\end{align*}
This proves \eqref{eq:principal_stratum_probabilities}.

For the always-$1$ stratum, monotonicity implies that $M(1)=1$ uniquely identifies $(M(1),M(0))=(1,1)$. Hence,
\begin{align*}
&\E[Y(1)\mid M(1)=1,M(0)=1,X=x]\\
&\quad=
\E[Y(1,1)\mid M(1)=1,M(0)=1,X=x]\\
&\quad=
\E[Y\mid A=1,M=1,S_M=1,X=x]\\
&\quad=
\E[Y\mid A=1,M=1,X=x]\\
&\quad=
\mu_{11}(x).
\end{align*}
The first equality follows from composition because $M(1)=1$ in this stratum. The second follows from marker-conditional exchangeability and consistency, and the third follows from outcome-marker conditional independence. Under $M(1)\leq M(0)$, $A=1$ and $M=1$ already imply $S_M=M(0)=1$, so the same result also follows directly from the observed treatment-arm regression.

For the never-$1$ stratum, $M(0)=0$ implies $M(1)=0$. Therefore,
\begin{align*}
&\E[Y(0)\mid M(1)=0,M(0)=0,X=x]\\
&\quad=
\E[Y(0,0)\mid S_M=0,X=x]\\
&\quad=
\E[Y\mid A=0,M=0,S_M=0,X=x]\\
&\quad=
\E[Y\mid A=0,M=0,X=x]\\
&\quad=
\mu_{00}(x).
\end{align*}
The first equality uses $S_M=M(0)=0$ and composition. The second follows from marker-conditional exchangeability and consistency, and the third follows from outcome-marker conditional independence.

The observed marker separates the two strata with $M(1)=0$ among treated individuals. If $(M(1),M(0))=(0,0)$, then $(A,M,S_M)=(1,0,0)$ under treatment. Hence,
\begin{align*}
&\E[Y(1)\mid M(1)=0,M(0)=0,X=x]\\
&\quad=
\E[Y(1,0)\mid M(1)=0,S_M=0,X=x]\\
&\quad=
\E[Y\mid A=1,M=0,S_M=0,X=x]\\
&\quad=
\mu_{0,1,0}(x)\\
&\quad=
\mu_{01}(x).
\end{align*}
The second equality follows from marker-conditional exchangeability and consistency, and the final equality follows from outcome-marker conditional independence.

Likewise, if $(M(1),M(0))=(0,1)$, then $(A,M,S_M)=(1,0,1)$ under treatment. Therefore,
\begin{align*}
&\E[Y(1)\mid M(1)=0,M(0)=1,X=x]\\
&\quad=
\E[Y(1,0)\mid M(1)=0,S_M=1,X=x]\\
&\quad=
\E[Y\mid A=1,M=0,S_M=1,X=x]\\
&\quad=
\mu_{0,1,1}(x)\\
&\quad=
\mu_{01}(x),
\end{align*}
where the second equality follows from marker-conditional exchangeability and consistency, and the last follows from outcome-marker conditional independence. These calculations establish \eqref{eq:principal_means_direct} and \eqref{eq:principal_means_marker}.

It remains to relate the two untreated means among individuals with $M(0)=1$. Define
\begin{align*}
\theta_{11}^{0}(x)
&=
\E[Y(0)\mid M(1)=1,M(0)=1,X=x],\\
\theta_{01}^{0}(x)
&=
\E[Y(0)\mid M(1)=0,M(0)=1,X=x].
\end{align*}
Among untreated individuals with $M=1$, consistency implies $M(0)=1$. Under monotonicity, these individuals belong either to the treatment-prevented stratum $(0,1)$ or to the always-$1$ stratum $(1,1)$. Consequently,
\begin{align*}
\mu_{10}(x)
&=
\E[Y\mid A=0,M=1,X=x]\\
&=
\theta_{01}^{0}(x)
P(M(1)=0,M(0)=1\mid A=0,M=1,X=x)\\
&\quad+
\theta_{11}^{0}(x)
P(M(1)=1,M(0)=1\mid A=0,M=1,X=x).
\end{align*}
By consistency, treatment exchangeability, and the principal-stratum probabilities derived above,
\begin{align*}
&P(M(1)=1,M(0)=1\mid A=0,M=1,X=x)\\
&\quad=
\frac{
P(M(1)=1,M(0)=1\mid X=x)
}{
P(M(0)=1\mid X=x)
}\\
&\quad=
\frac{p_{11}(x)}{p_{10}(x)},
\end{align*}
and
\begin{align*}
&P(M(1)=0,M(0)=1\mid A=0,M=1,X=x) \\
&\quad=
\frac{
P(M(1)=0,M(0)=1\mid X=x)
}{
P(M(0)=1\mid X=x)
}\\
&\quad=
\frac{p_{10}(x)-p_{11}(x)}{p_{10}(x)}.
\end{align*}
It follows that
\begin{align*}
\mu_{10}(x)
&=
\theta_{01}^{0}(x)
\frac{p_{10}(x)-p_{11}(x)}{p_{10}(x)}
+
\theta_{11}^{0}(x)
\frac{p_{11}(x)}{p_{10}(x)}.
\end{align*}
Multiplying by $p_{10}(x)$ and solving for $\theta_{01}^{0}(x)$ gives, provided $p_{10}(x)-p_{11}(x)>0$,
\begin{align*}
\theta_{01}^{0}(x)
&=
\frac{
p_{10}(x)\mu_{10}(x)
-
p_{11}(x)\theta_{11}^{0}(x)
}{
p_{10}(x)-p_{11}(x)
}.
\end{align*}
Equivalently,
\begin{align*}
&\E[Y(0)\mid M(1)=0,M(0)=1,X=x]\\
&\quad=
\frac{
p_{10}(x)\mu_{10}(x)
-
p_{11}(x)
\E[Y(0)\mid M(1)=1,M(0)=1,X=x]
}{
p_{10}(x)-p_{11}(x)
},
\end{align*}
which proves \eqref{eq:principal_prevented_mean}.

It remains to identify $\theta_{11}^{0}(x)$. Within the always-$1$ stratum, composition gives
\begin{align*}
\theta_{11}^{0}(x)
&=
\E[Y(0)\mid M(1)=1,M(0)=1,X=x]\\
&=
\E[Y(0,1)\mid M(1)=1,M(0)=1,X=x].
\end{align*}
Assumption~\ref{ass:ex_restriction} implies
\begin{align*}
\theta_{11}^{0}(x)
&=
\E[Y(1,1)\mid M(1)=1,M(0)=1,X=x].
\end{align*}
Because $M(1)=1$ in this stratum, another application of composition, followed by marker-conditional exchangeability and consistency, gives
\begin{align*}
\theta_{11}^{0}(x)
&=
\E[Y(1)\mid M(1)=1,M(0)=1,X=x]\\
&=
\E[Y\mid A=1,M=1,X=x]\\
&=
\mu_{11}(x).
\end{align*}
This proves \eqref{eq:principal_always_mean}. Substitution into \eqref{eq:principal_prevented_mean} gives
\begin{align*}
&\E[Y(0)\mid M(1)=0,M(0)=1,X=x]\\
&\quad=
\frac{
p_{10}(x)\mu_{10}(x)-p_{11}(x)\mu_{11}(x)
}{
p_{10}(x)-p_{11}(x)
},
\end{align*}
which proves \eqref{eq:principal_prevented_mean_identified}.

Finally, the conditional principal causal effects follow by subtracting the identified treatment-specific means within each stratum. For the always-$1$ stratum,
\begin{align*}
\tau_{11}(x)
&=
\E[Y(1)-Y(0)\mid M(1)=1,M(0)=1,X=x]\\
&=
\mu_{11}(x)-\mu_{11}(x)\\
&=0.
\end{align*}
For the never-$1$ stratum,
\begin{align*}
\tau_{00}(x)
&=
\E[Y(1)-Y(0)\mid M(1)=0,M(0)=0,X=x]\\
&=
\mu_{01}(x)-\mu_{00}(x).
\end{align*}
For the treatment-prevented stratum,
\begin{align*}
\tau_{01}(x)
&=
\E[Y(1)-Y(0)\mid M(1)=0,M(0)=1,X=x]\\
&=
\mu_{01}(x)
-
\frac{
p_{10}(x)\mu_{10}(x)-p_{11}(x)\mu_{11}(x)
}{
p_{10}(x)-p_{11}(x)
}.
\end{align*}
This proves \eqref{eq:identified_principal_effects}.

\hfill\(\square\)

\section{Generalizations to multiple mediators}
\label{app:multiple_mediators}

We illustrate how the susceptibility-marker construction extends to a setting with two ordered mediators, leaving a general treatment of multiple mediators to future work.

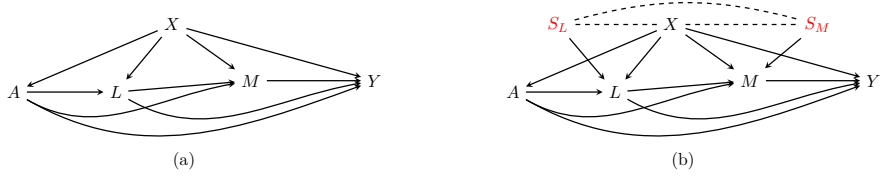
\begin{figure}[t] 
	\begin{center}
    \scalebox{0.6}{
    \begin{tikzpicture}[>=stealth, node distance=1.75cm]
        \tikzstyle{format} = [thick, circle, minimum sizS=1.0mm, inner sep=2pt]
        \tikzstyle{square} = [draw, thick, minimum size=4.5mm, inner sep=2pt]

    \begin{scope}[xshift=0.cm, yshift=0cm]
		\path[->, thick]
		
		node[] (a) {$A$}
		node[right of=a, xshift=0.5cm] (m1) {$L$}
        node[above right of=m1, yshift=0.25cm] (x) {$X$}
        node[below right of=x, xshift=0.5cm] (m2) {$M$}
        node[left of=x, xshift=-0.75cm] (s1) {}
        node[above right of=m2, xshift=0.25cm] (s2) {}
        node[below right of=s2, xshift=0.cm] (y) {$Y$}

        (x) edge[black] (a) 
        (x) edge[black] (m1) 
        (x) edge[black] (m2)
        (x) edge[black] (y)
		(a) edge[black] (m1) 
        (m1) edge[black] (m2) 
		(m2) edge[black] (y)
        (a) edge[black, out = -30, in = 190] (m2)
        (a) edge[black, out = -30, in = 200] (y)
        (m1) edge[black, out = -30, in = 190] (y)
        
        node[below of=m1, xshift=1.5cm, yshift=0.25cm] (t2) {(a)} ;
		
	\end{scope}
    \begin{scope}[xshift=11cm, yshift=0cm]
		\path[->, thick]
		
		node[] (a) {$A$}
		node[right of=a, xshift=0.5cm] (m1) {$L$}
        node[above right of=m1, yshift=0.25cm] (x) {$X$}
        node[below right of=x, xshift=0.5cm] (m2) {$M$}
        node[left of=x, xshift=-0.75cm] (s1) {\red{$S_L$}}
        node[above right of=m2, xshift=0.25cm] (s2) {\red{$S_M$}}
        node[below right of=s2, xshift=0.cm] (y) {$Y$}

        (x) edge[black] (a) 
        (x) edge[black] (m1) 
        (x) edge[black] (m2)
        (x) edge[black, dashed, -] (s1)
        (x) edge[black, dashed, -] (s2)
        (x) edge[black] (y)
		(a) edge[black] (m1) 
        (m1) edge[black] (m2) 
		(m2) edge[black] (y)
        (s1) edge[black] (m1)
        (s2) edge[black] (m2)
        (s1) edge[black, dashed, -, bend left=15] (s2)
        (a) edge[black, out = -30, in = 190] (m2)
        (a) edge[black, out = -30, in = 200] (y)
        (m1) edge[black, out = -30, in = 190] (y)
        
        node[below of=m1, xshift=1.5cm, yshift=0.25cm] (t3) {(b)} ;
		
	\end{scope}
    \end{tikzpicture}
    }
    \end{center}
    \vspace{-0.35cm}
    \caption{(a) Causal diagram with two ordered mediators $L$ and $M$; and (b) causal diagram with a candidate susceptibility marker $S_L$ for $L$ and a combined downstream marker $S_M$ summarizing the level-specific markers $\{S_{M,0},S_{M,1}\}$ for $M$. Under the corresponding necessity and sufficiency restrictions, each marker encodes the relevant mediator value under its reference treatment. Dashed edges denote marginal dependence that need not be causal.} 
    \label{fig:susceptibility_markers}
\end{figure}

Suppose that $A$ affects $Y$ through two ordered mediators, $L$ and $M$, where $L$ precedes $M$ in the causal order, as shown in Figure~\ref{fig:susceptibility_markers}(a). Consider a path-specific effect (PSE) through $M$, represented by the nested counterfactual $Y(a,L(a),M(a',L(a)))$. This is the potential outcome if treatment is set to $a$, the first mediator $L$ is set to its natural value under $A=a$, and the second mediator $M$ is set to the value it would take under treatment $A=a'$ and mediator level $L(a)$.

One natural decomposition of the total effect into a component through $M$ and a component not through $M$ is
\begin{align*}
\E[Y(1)-Y(0)]
&=
\left\{
\E[Y(1)]
-
\E[Y(1,L(1),M(0,L(1)))]
\right\}
\\
&\quad+
\left\{
\E[Y(1,L(1),M(0,L(1)))]
-
\E[Y(0)]
\right\}.
\end{align*}
Here, the first term captures the component through $M$, whereas the second captures the component not through $M$. As in the single-mediator setting, the total effect admits a second decomposition obtained by inserting the alternative nested counterfactual mean:
\begin{align*}
\E[Y(1)-Y(0)]
&=
\left\{
\E[Y(1)]
-
\E[Y(0,L(0),M(1,L(0)))]
\right\}
\\
&\quad+
\left\{
\E[Y(0,L(0),M(1,L(0)))]
-
\E[Y(0)]
\right\}.
\end{align*}
Under this decomposition, the first term is the component not through $M$, and the second is the component through $M$. The two decompositions differ in the treatment level at which the outcome and upstream mediator are evaluated and in the reference treatment level used for the downstream mediator. In the remainder of this section, we develop the first decomposition to illustrate the susceptibility-marker construction. The second follows by the same argument with the reference treatment levels reversed: the first uses markers satisfying $S_L=L(1)$ and $S_{M,l}=M(0,l)$, whereas the second would use markers satisfying $S_L=L(0)$ and $S_{M,l}=M(1,l)$. 

Identification of either decomposition under conventional potential-outcome models requires cross-world assumptions analogous to those invoked for natural direct and indirect effects. The following assumptions provide one sufficient identifying structure for the general nested mean $\E[Y(a,L(a),M(a',L(a)))]$. 

\begin{assumption}\label{ass:consistency2}
    \emph{Causal consistency:} If $A = a, L=l, M=m$, then $L(a) = L$, $M(a, l) = M$, and $Y(a, l, m) = Y$.
\end{assumption} 

\begin{assumption}\label{ass:positivity2}
    \emph{Positivity:} $P(A = a \mid X=x) > 0$, $P(L = l \mid A=a, X=x) > 0 $, and  $P(M = m \mid A=a, L=l, X=x) > 0$, for all possible levels of $a, l, m$, and all $x$ such that $P(X=x) > 0$.  
\end{assumption}

\begin{assumption}\label{ass:single_world_ignor2}
    \emph{Single-world ignorability:} $Y(a, l, m) \perp A, L, M \mid X$, $M(a, l) \perp A, L \mid X$, and $L(a) \perp A \mid X, \forall a, l, m$.
\end{assumption}

\begin{assumption}\label{ass:cross_world2}
\textit{Cross-world ignorability}: For all $l,m$ and $a\neq a'$,
\[
Y(a,l,m)
\perp
\{L(a),M(a',l)\}
\mid X
\quad 
\text{and} 
\quad
M(a',l)\perp L(a)\mid X.
\]
\end{assumption}
Under Assumptions~\ref{ass:consistency2}–\ref{ass:cross_world2}, 
\begin{align}
    \E[Y(a,L(a),M(a',L(a)))] = \int y \, dP(y \mid a, l, m, x)\, dP(m \mid a', l, x) \, dP(l \mid a, x) \, dP(x) \ . 
    \label{eq:id2}
\end{align}

Our goal is to show that susceptibility markers can instead represent this cross-world parameter as a mixture of single-world controlled counterfactual means. We then give conditions under which that mixture is identified. 

\subsection{Susceptibility markers in ordered-mediator setting} 

To generalize the single-mediator construction, we introduce baseline susceptibility markers illustrated in Figure~\ref{fig:susceptibility_markers}(b): an upstream marker $S_L$ for $L$ and a downstream family of markers $S_{M,l}$ for $M$, indexed by each possible value $l$ of the upstream mediator. The marker $S_L$ encodes susceptibility to activation of $L$ under the chosen reference treatment level $a$. For each $l\in\{0,1\}$, the marker $S_{M,l}$ encodes susceptibility to activation of $M$ under the reference treatment level $a'$ when $L$ is fixed to $l$. Under the necessity and sufficiency conditions introduced below, these markers will be shown to satisfy $S_L=L(a)$ and $S_{M,l}=M(a',l)$.

Together, the markers $\{S_L,S_{M,0},S_{M,1}\}$ are designed to recover the single-world potential mediators $L(a)$ and $M(a',l)$ that appear in the nested expression $M(a',L(a))$. Although $L(a)$ and $M(a',l)$ are each single-world counterfactuals, the composite term $M(a',L(a))$ is cross-world when $a\neq a'$ because it evaluates the $a'$-world mediator at the mediator level $L(a)$ arising under the $a$-world. This cross-world mediator then appears inside the nested outcome $Y(a,L(a),M(a',L(a)))$. Once the marker equalities are established, $L(a)$ can be replaced by $S_L$ and $M(a',L(a))$ by $S_{M,S_L}$, permitting the nested cross-world quantity to be represented as a mixture of single-world controlled counterfactuals indexed by baseline susceptibility markers.

The following assumptions extend the single-mediator necessity and sufficiency conditions to the ordered-mediators setting. The key idea mirrors the single-mediator case: introduce baseline markers whose restrictions imply the mediator values appearing in the nested counterfactual expression.

\begin{assumption}\label{ass:necessity2}
\textit{Necessity}: For each $b\in\{0,1\}$, all $x$, and all $l\in\{0,1\}$,
\begin{align*}
P(L(b)=1\mid S_L=0,X=x)=0,
\qquad
P(M(b,l)=1\mid S_{M,l}=0,X=x)=0.
\end{align*}
\end{assumption}

\begin{assumption}\label{ass:sufficiency2}
\textit{Sufficiency-under-reference-treatment}: For all $x$ and all $l\in\{0,1\}$,
\begin{align*}
P(L(a)=1\mid S_L=1,X=x)=1,
\qquad
P(M(a',l)=1\mid S_{M,l}=1,X=x)=1.
\end{align*}
\end{assumption}

Necessity requires that, if an individual is not susceptible according to the relevant marker, then the corresponding mediator cannot be activated under either treatment level. Sufficiency requires that, under the chosen reference treatment conditions, the presence of the marker guarantees activation of the corresponding mediator. 

\begin{lemma}\label{lem:sufficiency2}
Under Assumptions~\ref{ass:necessity2} and~\ref{ass:sufficiency2}, for every $l\in\{0,1\}$ 
\begin{align*}
S_L=L(a),
\qquad
S_{M,l}=M(a',l). 
\end{align*}
\end{lemma}

\noindent
The proof mirrors that of Lemma~\ref{lem:sufficiency}. Necessity ensures that the corresponding mediator is $0$ whenever its marker is $0$, while sufficiency ensures that the mediator is $1$ under the reference treatment whenever its marker is $1$. Together, these conditions establish equality of each marker with the corresponding potential mediator; see a proof in Appendix~\ref{app:proofs_sufficiency2}.

\begin{remark}\label{rem:monotonicity2}
\textbf{(Monotonicity implications).}
Assumptions~\ref{ass:necessity2} and~\ref{ass:sufficiency2} imply one-sided monotonicity conditions for the mediator potential outcomes. For the upstream mediator,
\[
L(a')\leq L(a).
\]
Indeed, if $S_L=0$, necessity implies $L(a')=L(a)=0$, whereas if $S_L=1$, sufficiency implies $L(a)=1$. Similarly, for each $l\in\{0,1\}$,
\[
M(a,l)\leq M(a',l).
\]
If $S_{M,l}=0$, necessity implies $M(a,l)=M(a',l)=0$, whereas if $S_{M,l}=1$, sufficiency implies $M(a',l)=1$. If the susceptibility markers are observed, consistency and adequate treatment and and upstream-mediator positivity make these necessity, sufficiency, and monotonicity conditions empirically falsifiable through their observed-data implications.
\end{remark}

Together, the markers $S_L$ and $\{S_{M,l}\}$ supply the single-world potential mediator values $L(a)$ and $M(a',l)$ appearing in the cross-world nested expression. In particular,
\[
M(a',L(a))
=
M(a',S_L)
=
S_{M,S_L}.
\]
The nested counterfactual mean can therefore be represented as a mixture of single-world controlled counterfactuals indexed by the baseline susceptibility markers.

\subsection{Rewriting the cross-world quantity as a single-world mixture}

We now show that $\E[Y(a,L(a),M(a',L(a)))]$ can be rewritten as a single-world mixture over the susceptibility markers $(S_L,S_{M,0},S_{M,1})$. Under Lemma~\ref{lem:sufficiency2},
\[
L(a)=S_L
\qquad\text{and}\qquad
M(a',L(a))=M(a',S_L)=S_{M,S_L}.
\]
Thus,
\begin{equation}\label{eq:Yalm}
\begin{aligned}
&\E[Y(a,L(a),M(a',L(a)))]
\\
&\quad=
\sum_{l=0}^1\sum_{m=0}^1
\E[
Y(a,l,m)
\mid
L(a)=l,M(a',L(a))=m
]
\,
P(L(a)=l,M(a',L(a))=m)
\\
&\quad=
\sum_{l=0}^1\sum_{m=0}^1
\E[
Y(a,l,m)
\mid
S_L=l,S_{M,l}=m
]
P(S_L=l,S_{M,l}=m).
\end{aligned}
\end{equation}

This representation mirrors \eqref{eq:Yam} in Section~\ref{sec:single_mediator}, but now involves an upstream marker and the downstream marker selected by its value. It contains only controlled potential outcomes $Y(a,l,m)$, which are single-world counterfactuals consistent with the FFRCISTG model, and probabilities of the corresponding susceptibility-marker strata.

As in the single-mediator case, the susceptibility-marker construction rewrites the cross-world estimand as a mixture of single-world controlled counterfactual means. When this representation is inserted into the total-effect decomposition, the resulting path-specific components can be expressed as weighted averages of susceptibility-stratum-specific contrasts between natural and controlled mediator regimes, with weights $P(S_L=l,S_{M,l}=m)$ determined by the joint distribution of $(S_L,S_{M,0},S_{M,1})$. Below, we provide an illustrative example.

At this stage, \eqref{eq:Yalm} establishes a single-world representation, not identification from observed data. In particular, necessity and sufficiency do not by themselves imply that the joint marker probabilities factor into mediator distributions from the two reference treatment arms. We return to identification after the example.

\subsubsection{Illustrative toy example}

To make the construction concrete, we present a toy example involving two ordered mediators. Let $A\in\{0,1\}$ represent assignment to a notional cognitive-training program. Let $L\in\{0,1\}$ denote an upstream mediator indicating whether a participant becomes \emph{engaged} during training sessions (e.g., maintains effort and attentional focus), and let $M\in\{0,1\}$ denote a downstream mediator corresponding to whether the participant subsequently undertakes \emph{additional practice outside the sessions} (e.g., completing optional exercises or using training apps). The outcome $Y$ is a generic measure of cognitive performance. This setting provides a natural motivation for the \emph{path-specific effect through $M$}. Investigators may ask how much of the effect of training on $Y$ operates through increases in self-initiated practice, as opposed to immediate effects of the training sessions or changes in engagement that do not lead to additional practice. The example is purely illustrative and serves only to clarify the role of susceptibility markers.

Suppose that engagement depends on stable characteristics such as motivation or self-regulatory capacity. Let the marker $S_L$ encode whether the participant possesses the characteristics required to become engaged. Under necessity, if $S_L=0$, the participant would not become engaged under either treatment level. Under sufficiency-under-treatment for $L$, if $S_L=1$, the participant would become engaged when assigned to the cognitive-training program. Combined, these conditions imply $ S_L=L(1)$. For the downstream mediator, introduce level-specific susceptibility markers $S_{M,l}$ for $l\in\{0,1\}$. The marker $S_{M,l}$ represents a participant's baseline capacity to initiate additional practice when the upstream mediator $L$ is fixed to $l$. Under necessity, if $S_{M,l}=0$, the participant would not undertake additional practice when $L=l$, regardless of treatment. Under sufficiency-under-control for $M$, if $S_{M,l}=1$, the participant would undertake additional practice under control when $L$ is fixed to $l$. Combined, these conditions imply $S_{M,l}=M(0,l)$. The mapping $l\mapsto S_{M,l}$ may vary across individuals. For example, some participants may undertake additional practice only when engaged, so that $(S_{M,0},S_{M,1})=(0,1)$, whereas others may practice regardless of engagement, so that $(S_{M,0},S_{M,1})=(1,1)$. Together, the markers $(S_L,S_{M,0},S_{M,1})$ permit the nested cross-world counterfactual $Y(1,L(1),M(0,L(1)))$ to be represented as a mixture of single-world controlled counterfactuals. Under the additional assumptions introduced below, the mean of this counterfactual can also be identified from observed data. 

\subsubsection{Representation of the path-specific effect through $M$}

The effect through $M$ is defined as $\E[Y(1)]-\E[Y(1,L(1),M(0,L(1)))]$. In this example, Lemma~\ref{lem:sufficiency2} gives $S_L=L(1)$ and $S_{M,l}=M(0,l)$ for $l\in\{0,1\}$. Based on these equalities, we can write
\begin{align*}
    &\hspace{-1cm}\E[Y(1)]-\E[Y(1,L(1),M(0,L(1)))] \\
    &=\sum_{l=0}^1\sum_{m=0}^1
    \E[Y(1)-Y(1,l,m)\mid S_L=l,S_{M,l}=m]
    P(S_L=l,S_{M,l}=m) \\
    &=\underbrace{\E[Y(1)-Y(1,1,1)\mid S_L=1,S_{M,1}=1]}_{\Delta^*_1}
    P(S_L=1,S_{M,1}=1) \\
    &\hspace{0.5cm}
    +\underbrace{\E[Y(1)-Y(1,1,0)\mid S_L=1,S_{M,1}=0]}_{\Delta^*_2}
    P(S_L=1,S_{M,1}=0) \\
    &\hspace{0.5cm}
    +\underbrace{\E[Y(1)-Y(1,0,1)\mid S_L=0,S_{M,0}=1]}_{\Delta^*_3}
    P(S_L=0,S_{M,0}=1) \\
    &\hspace{0.5cm}
    +\underbrace{\E[Y(1)-Y(1,0,0)\mid S_L=0,S_{M,0}=0]}_{\Delta^*_4}
    P(S_L=0,S_{M,0}=0).
\end{align*}

In strata with $S_{M,l}=0$, necessity implies that the downstream mediator cannot be activated when $L=l$, regardless of treatment, so that $M(1,l)=M(0,l)=0$. Because $S_L=l$ implies $L(1)=l$, composition gives
\[
Y(1)=Y(1,l,M(1,l))=Y(1,l,0)
\]
in these strata. Thus, the effect through $M$ is zero in these strata:
\[
\Delta^*_2=\Delta^*_4=0.
\]
It follows that the path-specific effect through $M$ is
\begin{align*}
&\E[Y(1)]-\E[Y(1,L(1),M(0,L(1)))]  
\\ 
&\hspace{3cm}= 
\Delta^*_1P(S_L=1,S_{M,1}=1)
+
\Delta^*_3P(S_L=0,S_{M,0}=1).
\end{align*}
The weights $P(S_L=l,S_{M,l}=m)$ are the prevalences of the corresponding susceptibility patterns in the population. Within the strata with $S_{M,l}=1$, units for whom $M(1,l)=1$ also contribute zero by composition. Therefore, a nonzero contribution can arise only from units for whom $M(0,l)=1$ and $M(1,l)=0$ at the relevant value of $l$.

\vspace{0.2cm}
\noindent\textbf{Interpretation of $\Delta^*_1$.}
The stratum $\{S_L=1,S_{M,1}=1\}$ consists of individuals who would have the upstream mediator $L$ active under treatment, $L(1)=1$, and whose marker indicates that the downstream mediator $M$ would be active under control when $L=1$, that is, $M(0,1)=1$. Within this group, $Y(1)$ denotes the outcome under $A=1$ with both mediators left at their natural values, $\{L(1),M(1,L(1))\}=\{1,M(1,1)\}$, whereas $Y(1,1,1)$ is the outcome under $A=1$ when we fix $(L,M)=(1,1)$. The contrast
\[
\Delta^*_1
=
\E[Y(1)-Y(1,1,1)\mid S_L=1,S_{M,1}=1]
\]
therefore measures, among units in the $\{S_L=1,S_{M,1}=1\}$ stratum, the change in outcome obtained by letting $M$ follow its natural value under treatment rather than fixing it to the reference active level it would take under control. Thus, $\Delta^*_1$ is a susceptibility-stratum-specific contrast between the natural downstream-mediator regime under treatment and the controlled downstream-mediator regime under which $M$ is fixed to its reference value. It should not be called a controlled indirect effect.

\vspace{0.2cm}
\noindent\textbf{Interpretation of $\Delta^*_3$.}
The stratum $\{S_L=0,S_{M,0}=1\}$ consists of individuals whose upstream mediator $L$ would remain inactive under either treatment level, $L(1)=L(0)=0$, and whose marker indicates that the downstream mediator $M$ would be active under control when $L=0$, that is, $M(0,0)=1$. For these units, $Y(1)$ is the outcome when both mediators take their natural values under $A=1$, here $\{L(1),M(1,L(1))\}=\{0,M(1,0)\},$ whereas $Y(1,0,1)$ is the outcome under $A=1$ with $(L,M)$ fixed at $(0,1)$. The contrast
\[
\Delta^*_3
=
\E[Y(1)-Y(1,0,1)\mid S_L=0,S_{M,0}=1]
\]
thus captures, among individuals whose upstream mediator remains inactive but whose downstream mediator would be active under control, the effect of allowing $M$ to take its natural value under $A=1$ rather than fixing it to the reference level $M(0,0)=1$. Like $\Delta^*_1$, $\Delta^*_3$ is a susceptibility-stratum-specific contrast between natural and controlled downstream-mediator regimes, not a controlled indirect effect.

\subsubsection{Representation of the path-specific effect *not* through $M$}

The effect not through $M$ is defined as
\begin{align*}
    &\hspace{-1cm}
    \E[Y(1,L(1),M(0,L(1)))]-\E[Y(0)]
    \\
    &=
    \sum_{l=0}^1\sum_{m=0}^1
    \E[
    Y(1,l,m)-Y(0)
    \mid
    S_L=l,S_{M,l}=m
    ]
    P(S_L=l,S_{M,l}=m)
    \\
    &=
    \underbrace{
    \E[
    Y(1,1,1)-Y(0)
    \mid
    S_L=1,S_{M,1}=1
    ]
    }_{\Delta_1}
    P(S_L=1,S_{M,1}=1)
    \\
    &\hspace{0.5cm}+
    \underbrace{
    \E[
    Y(1,1,0)-Y(0)
    \mid
    S_L=1,S_{M,1}=0
    ]
    }_{\Delta_2}
    P(S_L=1,S_{M,1}=0)
    \\
    &\hspace{0.5cm}+
    \underbrace{
    \E[
    Y(1,0,1)-Y(0)
    \mid
    S_L=0,S_{M,0}=1
    ]
    }_{\Delta_3}
    P(S_L=0,S_{M,0}=1)
    \\
    &\hspace{0.5cm}+
    \underbrace{
    \E[
    Y(1,0,0)-Y(0)
    \mid
    S_L=0,S_{M,0}=0
    ]
    }_{\Delta_4}
    P(S_L=0,S_{M,0}=0).
\end{align*}

Under the susceptibility markers $(S_L,S_{M,0},S_{M,1})$, this contrast decomposes into four susceptibility-stratum-specific contrasts $\Delta_1,\Delta_2,\Delta_3,\Delta_4$. For each stratum $k$,
\[
\Delta_k+\Delta_k^*
=
\E[Y(1)-Y(0)\mid\text{stratum }k],
\]
so that $\Delta_k$ and $\Delta_k^*$ give the respective ``not through $M$'' and ``through $M$'' components of the stratum-specific total effect. The precise interpretation of $\Delta_k$, however, depends on which mediator values are fixed by the corresponding marker stratum.

\vspace{0.2cm}
\noindent\textbf{Interpretation of $\Delta_1$.}
Individuals in the stratum $\{S_L=1,S_{M,1}=1\}$ would have the upstream mediator active under treatment, $L(1)=1$, and their downstream marker indicates that the downstream mediator would be active under control when $L=1$, that is, $M(0,1)=1$. For this group,
\[
\Delta_1
=
\E[
Y(1,1,1)-Y(0)
\mid
S_L=1,S_{M,1}=1
]
\]
compares the outcome under $A=1$ with $(L,M)$ fixed at $(1,1)$ with the outcome under $A=0$ when both mediators are left at their natural values.

Although $S_L=1$ establishes $L(1)=1$, it does not necessarily determine $L(0)$. Consequently, the two counterfactual outcomes in $\Delta_1$ do not generally hold the mediators fixed at the same levels. Thus, $\Delta_1$ is not, in general, a conditional controlled direct effect. Rather, it is a susceptibility-stratum-specific contrast between a controlled mediator regime under treatment and the natural mediator regime under control. Together, $\Delta_1$ and $\Delta_1^*$ decompose the total effect in this stratum into the components not through and through $M$, respectively.

\vspace{0.2cm}
\noindent\textbf{Interpretation of $\Delta_2$.}
Individuals in the stratum $\{S_L=1,S_{M,1}=0\}$ would have the upstream mediator active under treatment, $L(1)=1$, but are not susceptible to downstream activation when $L=1$. Necessity therefore implies $M(1,1)=M(0,1)=0$. For this group,
\[
\Delta_2
=
\E[
Y(1,1,0)-Y(0)
\mid
S_L=1,S_{M,1}=0
].
\]
Because $L(1)=1$ and $M(1,1)=0$, composition gives $Y(1)=Y(1,1,0)$ within this stratum. Hence,
\[
\Delta_2
=
\E[
Y(1)-Y(0)
\mid
S_L=1,S_{M,1}=0
],
\]
so $\Delta_2$ equals the entire stratum-specific total effect and $\Delta_2^*=0$.

This conclusion does not imply that the downstream mediator is structurally inactive under the natural control regime. In particular, $S_L=1$ does not determine $L(0)$. If $L(0)=0$, the natural downstream mediator under control is $M(0,0)$, which is not restricted by $S_{M,1}=0$. Thus, $\Delta_2$ equals the stratum-specific total effect algebraically because the through-$M$ component of this decomposition is zero, but it should not generally be interpreted as a controlled direct effect or as an effect that necessarily operates entirely outside the downstream mediator under both natural treatment regimes.

\vspace{0.2cm}
\noindent\textbf{Interpretation of $\Delta_3$.}
Individuals in the stratum $\{S_L=0,S_{M,0}=1\}$ would have the upstream mediator remain inactive under both treatment levels, $L(1)=L(0)=0$, and their downstream marker indicates that they would have $M(0,0)=1$ under control when $L=0$. For this group,
\[
\Delta_3
=
\E[
Y(1,0,1)-Y(0)
\mid
S_L=0,S_{M,0}=1
].
\]
Because $L(0)=0$ and $M(0,0)=1$, composition gives $Y(0)=Y(0,0,1)$. Therefore,
\[
\Delta_3
=
\E[
Y(1,0,1)-Y(0,0,1)
\mid
S_L=0,S_{M,0}=1
],
\]
which is a conditional controlled direct effect with the mediators fixed at $(L,M)=(0,1)$. The pair $(\Delta_3,\Delta_3^*)$ decomposes the stratum-specific total effect into a controlled direct-effect component not through $M$ and a susceptibility-stratum-specific natural-versus-controlled mediator-regime contrast through $M$.

\vspace{0.2cm}
\noindent\textbf{Interpretation of $\Delta_4$.}
Individuals in the stratum $\{S_L=0,S_{M,0}=0\}$ would have neither mediator activated under either treatment level when $L=0$. In particular, necessity implies $L(1)=L(0)=0$ and $M(1,0)=M(0,0)=0$. For this group,
\[
\Delta_4
=
\E[
Y(1,0,0)-Y(0)
\mid
S_L=0,S_{M,0}=0
].
\]
By composition, $Y(0)=Y(0,0,0)$ and $Y(1)=Y(1,0,0)$. It follows that
\[
\Delta_4
=
\E[
Y(1,0,0)-Y(0,0,0)
\mid
S_L=0,S_{M,0}=0
],
\]
which is a conditional controlled direct effect with both mediators fixed at zero. Because $\Delta_4^*=0$, $\Delta_4$ also equals the entire stratum-specific total effect. In this stratum, treatment does not change either mediator, so the total effect operates through pathways other than changes in $L$ or $M$. 

\medskip
In summary, $\Delta_1$ is a controlled-versus-natural mediator-regime contrast and is not generally a conditional controlled direct effect. The contrast $\Delta_2$ equals the stratum-specific total effect because $\Delta_2^*=0$, but it is also not generally a controlled direct effect because the natural mediator values under control need not equal $(1,0)$. By contrast, $\Delta_3$ and $\Delta_4$ reduce to conditional controlled direct effects because, in their respective marker strata, composition allows $Y(0)$ to be written as $Y(0,0,1)$ and $Y(0,0,0)$, matching the controlled mediator levels in the corresponding treatment counterfactuals. In every stratum, the pair $(\Delta_k,\Delta_k^*)$ provides the exact ``not through $M$'' and ``through $M$'' decomposition of the stratum-specific total effect.

\subsection{Identification of the single-world mixture}

Representation of the cross-world counterfactual as the mixture in \eqref{eq:Yalm} does not require the susceptibility markers to be observed. Identification from observed data requires conditions linking the controlled counterfactual means within marker strata to observed outcome regressions and conditions identifying the joint distribution of the marker strata.

\begin{assumption}\label{ass:blinding2}
\textit{Marker blinding and sequential marker exchangeability}: For every $l\in\{0,1\}$,
\[
(S_L,S_{M,l})\perp A\mid X
\quad
\text{and}
\quad 
S_{M,l}\perp L\mid A=a',X,S_L.
\]
\end{assumption}

\begin{assumption}\label{ass:marker_positivity2}
\textit{Marker-stratum positivity}: For all $l,m\in\{0,1\}$ and all $x$ such that $P(S_L=l,S_{M,l}=m\mid X=x)>0,$ we have $P(A=a,L=l,M=m \mid X=x,S_L=l,S_{M,l}=m)>0.$ In addition, for all $l$ and $x$ such that $P(S_L=l\mid X=x)>0$, $P(A=a',L=l\mid X=x,S_L=l)>0$. 
\end{assumption}

\begin{assumption}\label{ass:marker_exch2}
\textit{Marker-conditional single-world exchangeability}: For all $l,m\in\{0,1\}$,
\begin{align*}
Y(a,l,m)
&\perp A
\mid X,S_L,S_{M,l},
\\
Y(a,l,m)
&\perp L
\mid A=a,X,S_L,S_{M,l},
\\
Y(a,l,m)
&\perp M
\mid A=a,L=l,X,S_L,S_{M,l}.
\end{align*}
\end{assumption}

\begin{assumption}\label{ass:indep2}
\textit{Mediator- and outcome-marker conditional independence}: For every $l\in\{0,1\}$,
\[
M\perp S_L\mid A,L,X
\quad
\text{and}
\quad 
Y\perp (S_L,S_{M,l})\mid A,L,M,X.
\]
\end{assumption}

Assumption~\ref{ass:blinding2} states that treatment assignment does not select individuals according to their susceptibility markers and that, conditional on $(X,S_L)$, the observed value of the upstream mediator under the reference treatment $a'$ does not select individuals according to the downstream marker $S_{M,l}$. Assumption~\ref{ass:marker_exch2} provides the single-world exchangeability conditions needed to identify controlled counterfactual means within marker strata. Assumption~\ref{ass:indep2} allows the observed mediator and outcome regressions to be collapsed over the latent markers.

Under these conditions, the joint marker distribution is identified even when the markers are not observed. By Lemma~\ref{lem:sufficiency2}, marker blinding, sequential marker exchangeability, consistency, and Assumption~\ref{ass:indep2},
\begin{align}
&P(S_L=l,S_{M,l}=m\mid X=x)
\notag\\
&\quad=
P(L=l\mid A=a,X=x)
P(M=m\mid A=a',L=l,X=x).
\label{eq:joint_marker_id2}
\end{align}
Consequently, the corresponding marker-stratum probabilities are identified by
\begin{align}
&P(S_L=l,S_{M,l}=m)
\notag\\
&\quad=
\int
P(L=l\mid A=a,X=x)
P(M=m\mid A=a',L=l,X=x)\,dP(x),
\label{eq:marker_prevalence2}
\end{align}
and, provided this probability is positive, the covariate distribution within the marker stratum is identified by
\begin{align}
&dP(x\mid S_L=l,S_{M,l}=m)
\notag\\
&\quad=
\frac{
P(L=l\mid A=a,X=x)
P(M=m\mid A=a',L=l,X=x)\,dP(x)
}{
\displaystyle
\int
P(L=l\mid A=a,X=x')
P(M=m\mid A=a',L=l,X=x')\,dP(x')
}.
\label{eq:marker_covariate_distribution2}
\end{align}

We formulate the identification result in the following theorem.

\begin{theorem}\label{thm:id_single_world_2}
Under Assumptions~\ref{ass:consistency2}, \ref{ass:positivity2},
\ref{ass:necessity2}, \ref{ass:sufficiency2}, and
\ref{ass:blinding2}--\ref{ass:indep2},
\begin{align}
&\E[Y(a,L(a),M(a',L(a)))]
\notag\\
&\quad=
\int
\sum_{l=0}^1\sum_{m=0}^1
\E[Y\mid A=a,L=l,M=m,X=x]
\notag\\
&\qquad\qquad\times
P(M=m\mid A=a',L=l,X=x)
P(L=l\mid A=a,X=x)\,dP(x).
\label{eq:id_marker_multiple}
\end{align}
\end{theorem}

See Appendix~\ref{app:proofs_id_single_world_2} for a proof.

Thus, the susceptibility markers need not be observed for identification. The assumptions identify both the controlled counterfactual means within marker strata and the distributions of those strata from the observed data. The resulting functional in \eqref{eq:id_marker_multiple} agrees with the conventional functional in \eqref{eq:id2}, but it is obtained through single-world assumptions involving the susceptibility markers rather than the cross-world independence condition in Assumption~\ref{ass:cross_world2}. When the markers are latent, these conditions remain substantive and generally untestable restrictions.

\subsection{Proof of Lemma~\ref{lem:sufficiency2}}
\label{app:proofs_sufficiency2}

Consider first the upstream mediator. Because $L(a)$ and $S_L$ are binary, for almost every $x$,
\begin{align*}
P(L(a)=S_L\mid X=x)
&=
P(L(a)=0,S_L=0\mid X=x)
+
P(L(a)=1,S_L=1\mid X=x)
\\
&\quad=
P(L(a)=0\mid S_L=0,X=x)
P(S_L=0\mid X=x)
\\
&\qquad+
P(L(a)=1\mid S_L=1,X=x)
P(S_L=1\mid X=x).
\end{align*}
By Assumption~\ref{ass:necessity2}, $P(L(a)=0\mid S_L=0,X=x)=1$, and by Assumption~\ref{ass:sufficiency2}, $P(L(a)=1\mid S_L=1,X=x)=1$. Therefore,
\begin{align*}
P(L(a)=S_L\mid X=x)
&=
P(S_L=0\mid X=x)
+
P(S_L=1\mid X=x)
=1,
\end{align*}
which establishes $L(a)=S_L$ almost surely.

Next, fix $l\in\{0,1\}$. Because $M(a',l)$ and $S_{M,l}$ are binary, for almost every $x$,
\begin{align*}
&P(M(a',l)=S_{M,l}\mid X=x)
\\
&\quad=
P(M(a',l)=0,S_{M,l}=0\mid X=x)
+
P(M(a',l)=1,S_{M,l}=1\mid X=x)
\\
&\quad=
P(M(a',l)=0\mid S_{M,l}=0,X=x)
P(S_{M,l}=0\mid X=x)
\\
&\qquad+
P(M(a',l)=1\mid S_{M,l}=1,X=x)
P(S_{M,l}=1\mid X=x).
\end{align*}
By Assumption~\ref{ass:necessity2}, $P(M(a',l)=0\mid S_{M,l}=0,X=x)=1$, and by Assumption~\ref{ass:sufficiency2}, $P(M(a',l)=1\mid S_{M,l}=1,X=x)=1$. Consequently,
\begin{align*}
P(M(a',l)=S_{M,l}\mid X=x)
&=
P(S_{M,l}=0\mid X=x)
+
P(S_{M,l}=1\mid X=x)
=1.
\end{align*}
Thus, $M(a',l)=S_{M,l}$ almost surely for every $l\in\{0,1\}$, which completes the proof.

\subsection{Proof of Theorem~\ref{thm:id_single_world_2}}
\label{app:proofs_id_single_world_2}

By the definition of the nested counterfactual and iterated expectation,
\begin{align*}
\E[Y(a,L(a),M(a',L(a)))]
&=
\int
\sum_{l=0}^1\sum_{m=0}^1
\E[
Y(a,l,m)
\mid
L(a)=l,M(a',L(a))=m,X=x
]
\\
&\qquad\qquad\times
P\{L(a)=l,M(a',L(a))=m\mid X=x\}\,dP(x).
\end{align*}

By Lemma~\ref{lem:sufficiency2}, $L(a)=S_L$ and $M(a',L(a))=M(a',S_L)=S_{M,S_L}$. Consequently, on the event $\{S_L=l\}$,
$M(a',L(a))=S_{M,l}$, and hence
\begin{align*}
\E[Y(a,L(a),M(a',L(a)))]
&=
\int
\sum_{l=0}^1\sum_{m=0}^1
\E[
Y(a,l,m)
\mid
S_L=l,S_{M,l}=m,X=x
]
\\
&\qquad\qquad\times
P(S_L=l,S_{M,l}=m\mid X=x)\,dP(x).
\end{align*}

We first identify the controlled counterfactual mean within each marker stratum. For every $(l,m,x)$ such that $P(S_L=l,S_{M,l}=m\mid X=x)>0,$ Assumptions~\ref{ass:marker_positivity2} and~\ref{ass:marker_exch2} imply
\begin{align*}
&\E[
Y(a,l,m)
\mid
S_L=l,S_{M,l}=m,X=x
]
\\
&\quad=
\E[
Y(a,l,m)
\mid
A=a,S_L=l,S_{M,l}=m,X=x
]
\\
&\quad=
\E[
Y(a,l,m)
\mid
A=a,L=l,S_L=l,S_{M,l}=m,X=x
]
\\
&\quad=
\E[
Y(a,l,m)
\mid
A=a,L=l,M=m,S_L=l,S_{M,l}=m,X=x
].
\end{align*}
These equalities follow, respectively, from treatment exchangeability, exchangeability for the upstream mediator, and exchangeability for the downstream mediator.

By consistency,
\begin{align*}
&\E[
Y(a,l,m)
\mid
A=a,L=l,M=m,S_L=l,S_{M,l}=m,X=x
]
\\
&\quad=
\E[
Y
\mid
A=a,L=l,M=m,S_L=l,S_{M,l}=m,X=x
].
\end{align*}
The outcome-marker conditional independence in Assumption~\ref{ass:indep2} then gives
\begin{align*}
&\E[
Y
\mid
A=a,L=l,M=m,S_L=l,S_{M,l}=m,X=x
]
\\
&\quad=
\E[
Y
\mid
A=a,L=l,M=m,X=x
].
\end{align*}
Therefore,
\begin{align}
&\E[
Y(a,l,m)
\mid
S_L=l,S_{M,l}=m,X=x
]
\notag\\
&\quad=
\E[
Y
\mid
A=a,L=l,M=m,X=x
].
\label{eq:proof_controlled_mean2}
\end{align}
It remains to identify the joint marker distribution. By probability factorization,
\begin{align*}
&P(S_L=l,S_{M,l}=m\mid X=x)
=
P(S_L=l\mid X=x)
P(S_{M,l}=m\mid S_L=l,X=x).
\end{align*}

For the first factor, marker blinding gives
\begin{align*}
P(S_L=l\mid X=x)
&=
P(S_L=l\mid A=a,X=x)
=
P(L=l\mid A=a,X=x),
\end{align*}
where the second equality follows because Lemma~\ref{lem:sufficiency2} gives $S_L=L(a)$ and, under $A=a$, consistency gives $L=L(a)$.

For the second factor, marker blinding and sequential marker exchangeability give
\begin{align*}
&P(S_{M,l}=m\mid S_L=l,X=x)
\\
&\quad=
P(S_{M,l}=m\mid A=a',S_L=l,X=x)
\\
&\quad=
P(S_{M,l}=m
\mid A=a',L=l,S_L=l,X=x).
\end{align*}
By Lemma~\ref{lem:sufficiency2},
$S_{M,l}=M(a',l)$. Under $A=a'$ and $L=l$, consistency therefore gives $M=M(a',l)=S_{M,l}.$ It follows that
\begin{align*}
&P(S_{M,l}=m
\mid A=a',L=l,S_L=l,X=x)
=
P(M=m
\mid A=a',L=l,S_L=l,X=x).
\end{align*}
The mediator-marker conditional independence in Assumption~\ref{ass:indep2} then yields
\begin{align*}
&P(M=m
\mid A=a',L=l,S_L=l,X=x)
=
P(M=m\mid A=a',L=l,X=x).
\end{align*}
Combining these equalities,
\begin{align}
&P(S_L=l,S_{M,l}=m\mid X=x)
=
P(L=l\mid A=a,X=x)
P(M=m\mid A=a',L=l,X=x),
\label{eq:proof_joint_marker2}
\end{align}
which proves \eqref{eq:joint_marker_id2}. Integrating \eqref{eq:proof_joint_marker2} over $P(x)$ gives \eqref{eq:marker_prevalence2}, and Bayes' rule gives \eqref{eq:marker_covariate_distribution2}.

Finally, substituting \eqref{eq:proof_controlled_mean2} and \eqref{eq:proof_joint_marker2} into the marker-mixture representation gives
\begin{align*}
&\E[Y(a,L(a),M(a',L(a)))]
\\
&\quad=
\int
\sum_{l=0}^1\sum_{m=0}^1
\E[Y\mid A=a,L=l,M=m,X=x]
\\
&\qquad\qquad\times
P(M=m\mid A=a',L=l,X=x)
P(L=l\mid A=a,X=x)\,dP(x),
\end{align*}
which proves \eqref{eq:id_marker_multiple}.

\section{Implications for exposure-induced confounding}
\label{app:exposure_induced_confounding}

Consider a longitudinal mediation structure with exposure-induced confounding, where $A\in\{0,1\}$ is the exposure, $L$ is an exposure-induced mediator--outcome confounder, $M$ is a downstream mediator, $Y$ is the outcome, and $X$ denotes baseline covariates; see Figure~\ref{fig:susceptibility_markers}(a).

A path-specific decomposition that isolates pathways through $M$ can be formulated using nested counterfactuals of the form
\[
Y(a,L(a),M(a',L(a'))),
\qquad a\neq a'.
\]
This counterfactual sets $A$ to $a$ in the outcome model, sets $L$ to its natural value under $A=a$, and sets $M$ to the value it would take under $A=a'$ with $L$ set to its natural value under $A=a'$. The corresponding decomposition of the total effect is
\begin{align*}
\E[Y(1)-Y(0)]
&=
\left\{
\E[Y(1)]
-
\E[Y(1,L(1),M(0,L(0)))]
\right\}
\\
&\quad+
\left\{
\E[Y(1,L(1),M(0,L(0)))]
-
\E[Y(0)]
\right\},
\end{align*}
where the first term captures the component through $M$ and the second captures the component not through $M$.

Even under an NPSEM-IE, cross-world parameters such as
\[
\E[Y(a,L(a),M(a',L(a')))]
\]
are generally not identified from observed data in the presence of exposure-induced confounding because $L$ is a \emph{recanting witness} for pathways involving $M$ \citep{chen05ijcai}. Under the NPSEM-IE independence restrictions, consistency, exchangeability, and positivity, the parameter can be written as
\begin{align}
\E[Y(a,L(a),M(a',L(a')))]
&=
\int
\sum_{l=0}^1\sum_{l'=0}^1\sum_{m=0}^1
\E[Y\mid A=a,L=l,M=m,X=x]
\notag\\
&\qquad\qquad\times
P(M=m\mid A=a',L=l',X=x)
\notag\\
&\qquad\qquad\times
P(L(a)=l,L(a')=l'\mid X=x)\,dP(x).
\label{eq:recanting_functional}
\end{align}

The NPSEM-IE independence restrictions justify the factorization in \eqref{eq:recanting_functional}: conditional on $X$, the controlled outcome $Y(a,l,m)$, the downstream mediator $M(a',l')$, and the pair $\{L(a),L(a')\}$ are generated by independent structural disturbances. Exchangeability and consistency then link the first two counterfactual distributions to the corresponding observed outcome and mediator regressions. The remaining component, $P(L(a)=l,L(a')=l'\mid X=x),$ is the conditional joint distribution of the potential outcomes of the recanting witness under conflicting exposure values. The observed data identify its two marginal distributions but not, without further restrictions, their joint distribution.

For binary $L$, \citet{tchetgen2014identification} showed that this joint distribution becomes identified under the one-sided monotonicity condition
\[
L(a')\leq L(a).
\]
Specifically,
\begin{align}
&P(L(a)=l,L(a')=l'\mid X=x)
\notag\\
&\hspace{1.5cm}=
\begin{cases}
P(L=1\mid A=a',X=x),
& l=1,\ l'=1,
\\[4pt]
P(L=1\mid A=a,X=x)
-
P(L=1\mid A=a',X=x),
& l=1,\ l'=0,
\\[4pt]
0,
& l=0,\ l'=1,
\\[4pt]
P(L=0\mid A=a,X=x),
& l=0,\ l'=0.
\end{cases}
\label{eq:monotone_joint_L}
\end{align}
Under the NPSEM-IE restrictions used in \eqref{eq:recanting_functional}, substituting \eqref{eq:monotone_joint_L} identifies the path-specific counterfactual mean.

\paragraph{Connection to the susceptibility marker.}
The susceptibility-marker construction makes the source of this monotonicity restriction explicit. Consider an upstream susceptibility marker $S_L$ satisfying the necessity and sufficiency-under-reference-treatment conditions for $L$ introduced in Section~\ref{app:multiple_mediators}. If $a$ is the reference treatment level, Lemma~\ref{lem:sufficiency2} gives
\[
S_L=L(a).
\]
Moreover, necessity and sufficiency together imply
\[
L(a')\leq L(a);
\]
see Remark~\ref{rem:monotonicity2}. Thus, the susceptibility-marker conditions provide a substantive structure under which the monotonicity restriction used in \eqref{eq:monotone_joint_L} holds.

When $S_L$ is observed, it also directly reveals one member of the pair $\{L(a),L(a')\}$. Under consistency and treatment exchangeability,
\[
\{S_L,L(a')\}\perp A\mid X,
\]
we have
\begin{align}
&P(L(a)=l,L(a')=l'\mid X=x)
=
P(S_L=l,L=l'\mid A=a',X=x).
\label{eq:joint_L_marker}
\end{align}
Thus, among individuals assigned $A=a'$, the observed pair $(S_L,L)$ reveals the pair $(L(a),L(a'))$. The joint distribution required in \eqref{eq:recanting_functional} can therefore be recovered directly from the observed marker and mediator distribution. Under the necessity and sufficiency restrictions, this observed joint distribution necessarily satisfies the monotonicity restrictions in \eqref{eq:monotone_joint_L}.

\paragraph{What the susceptibility marker does and does not identify.}
The distinction from the ordered-mediator setting considered in Section~\ref{app:multiple_mediators} is important. There, the nested counterfactual
\[
Y(a,L(a),M(a',L(a)))
\]
contains the same upstream potential mediator $L(a)$ in both occurrences, so that the markers $S_L=L(a)$ and $S_{M,l}=M(a',l)$ encode all mediator values appearing in the nested expression. By contrast, the counterfactual considered here,
\[
Y(a,L(a),M(a',L(a'))),
\]
contains both $L(a)$ and $L(a')$. The marker $S_L=L(a)$ reveals only the first of these values.

To see what remains unidentified outside the NPSEM-IE, iterated expectation gives
\begin{align*}
&\E[Y(a,L(a),M(a',L(a')))]
\\
&\quad=
\int
\sum_{l=0}^1\sum_{l'=0}^1\sum_{m=0}^1
\E[
Y(a,l,m)
\mid
L(a)=l,L(a')=l',
M(a',l')=m,X=x
]
\\
&\qquad\qquad\times
P\{
M(a',l')=m
\mid
L(a)=l,L(a')=l',X=x
\}
\\
&\qquad\qquad\times
P\{
L(a)=l,L(a')=l'
\mid X=x
\}\,dP(x).
\end{align*}
An observed marker satisfying $S_L=L(a)$ identifies the final factor through \eqref{eq:joint_L_marker}. It does not, by itself, identify the first two factors, which involve the controlled outcome and downstream mediator within strata jointly defined by $L(a)$ and $L(a')$.

Under the NPSEM-IE, the relevant independence restrictions remove this dependence on the joint response type of $L$, yielding the factorization in \eqref{eq:recanting_functional}. Those restrictions do not follow from the susceptibility-marker conditions. Likewise, adding the downstream monotonicity restriction
\[
M(a,l)\leq M(a',l)
\]
would restrict the possible response types of $M$ but would not identify either
\[
\E[
Y(a,l,m)
\mid
L(a)=l,L(a')=l',
M(a',l')=m,X=x
]
\]
or
\[
P\{
M(a',l')=m
\mid
L(a)=l,L(a')=l',X=x
\}.
\]
Monotonicity determines which response types are possible, but not their associations with the controlled outcome or with the response type of $L$.

Consequently, the susceptibility marker $S_L$ alone does not yield nonparametric identification of $\E[Y(a,L(a),M(a',L(a')))]$ under a single-world model that does not supply the NPSEM-IE cross-world independencies. Full identification would require additional assumptions identifying the two remaining stratum-specific quantities above. Such assumptions might be formulated using scientifically justified baseline variables together with appropriate marker-conditional exchangeability and conditional-independence conditions, but they are not consequences of the susceptibility-marker construction developed here. Alternatively, one may retain the NPSEM-IE independence restrictions used in \eqref{eq:recanting_functional} or impose another identifying structure appropriate to the scientific setting.

Thus, the susceptibility-marker restrictions provide a substantive basis for the monotonicity condition used to recover the recanting-witness distribution under the NPSEM-IE, but they do not replace the additional cross-world restrictions needed to identify the remaining stratum-specific outcome and downstream-mediator distributions.

\end{document}